\documentclass[12pt,letterpaper,a4paper]{article}
\usepackage[includeheadfoot,
            marginratio={1:1,2:3},
            width=475pt,
            height=710pt,]{geometry}
\usepackage{amsmath}
\usepackage{amsfonts}
\usepackage{amssymb}
\usepackage{graphicx}
\usepackage{cancel}

\usepackage{empheq}
\usepackage{color}
\usepackage{hyperref}

\usepackage{float}
\restylefloat{table}
\restylefloat{figure}

\newcommand{\nc}{\newcommand}
\nc{\beq}{\begin{equation}}
\nc{\eeq}{\end{equation}}
\nc{\bea}{\begin{eqnarray}}
\nc{\eea}{\end{eqnarray}}
\def\IZ{\mathbb{Z}}

\def\ov{\overline}

\begin{document}

\vspace{1.5cm}
\begin{center}
{\LARGE
On modular completion of generalized flux orbits}
\vspace{0.4cm}
\end{center}

\vspace{0.35cm}
\begin{center}
Pramod Shukla \footnote{Email: pkshukla@to.infn.it}
\end{center}

\vspace{0.1cm}
\begin{center}
{
%$^{\dag}$Department of Physics, Robeson Hall,  Virginia Tech,
%\vskip0.02cm
%850 West Campus Drive, Blacksburg, VA 24061, USA \\
%\vskip0.4cm
Universit\'a di Torino, Dipartimento di Fisica and I.N.F.N.-sezione di Torino
\vskip0.02cm
Via P. Giuria 1, I-10125 Torino, Italy \footnote{From October 1, 2015, the address has been changed to {\it ICTP, Strada Costiera 11, Trieste 34014, Italy} along with the new Email: shukla.pramod@ictp.it. }
}
\end{center}

\vspace{1cm}

%%%%%%%%%%%%%%%%%%%%%%%%%%%%%%%%%%%%%%%%%%%%%%%
%%%%%%%%%%%%%%%%%%%%%%%%%%%%%%%%%%%%%%%%%%%%%%%
%%%%%%%%%%%%%%%%%%%%%%%%%%%%%%%%%%%%%%%%%%%%%%%
%%%%%%%%%%%%%%%%%%%%%%%%%%%%%%%%%%%%%%%%%%%%%%%
%%%%%%%%%%%%%%%%%%%%%%%%%%%%%%%%%%%%%%%%%%%%%%%
%%%%%%%%%%%%%%%%%%%%%%%%%%%%%%%%%%%%%%%%%%%%%%%
%%%%%%%%%%%%%%%%%%%%%%%%%%%%%%%%%%%%%%%%%%%%%%%
%%%%%%%%%%%%%%%%%%%%%%%%%%%%%%%%%%%%%%%%%%%%%%%

\begin{abstract}
In the context of type IIB orientifold compactification with the presence of (non-)geometric fluxes, we conjecture a modular completed version of the generalized flux-orbits of various NS-NS and RR fluxes. Subsequently, considering two explicit examples with frozen complex structure moduli, we illustrate the utility of these new flux orbits in a very compact rearrangement of the four dimensional effective scalar potential.
\end{abstract}

\clearpage

\tableofcontents

%%%%%%%%%%%%%%%%%%%%%%%%%%%%%%%%%%%%%%%%%%%%%%%
%%%%%%%%%%%%%%%%%%%%%%%%%%%%%%%%%%%%%%%%%%%%%%%
%%%%%%%%%%%%%%%%%%%%%%%%%%%%%%%%%%%%%%%%%%%%%%%
%%%%%%%%%%%%%%%%%%%%%%%%%%%%%%%%%%%%%%%%%%%%%%%
%%%%%%%%%%%%%%%%%%%%%%%%%%%%%%%%%%%%%%%%%%%%%%%
%%%%%%%%%%%%%%%%%%%%%%%%%%%%%%%%%%%%%%%%%%%%%%%
%%%%%%%%%%%%%%%%%%%%%%%%%%%%%%%%%%%%%%%%%%%%%%%
%%%%%%%%%%%%%%%%%%%%%%%%%%%%%%%%%%%%%%%%%%%%%%%

%\newpage

\section{Introduction}
The interesting connections between gauged supergravities and string compactifications have attracted a lot of attention in recent years \cite{Derendinger:2004jn, Derendinger:2005ph, Shelton:2005cf,Dall'Agata:2009gv,Aldazabal:2011yz,Aldazabal:2011nj,Geissbuhler:2011mx,Grana:2012rr,Dibitetto:2012rk,Aldazabal:2006up,Hull:2004in, Andriot:2013xca, Blair:2014zba, Andriot:2012an, Andriot:2011uh, Andriot:2012wx,Andriot:2014qla,Kachru:2002sk, Hellerman:2002ax,Dabholkar:2002sy}. In the framework of type II orientifolds, a successive application of T-duality on the three form $H$-flux results in a chain of geometric and non-geometric fluxes as
\bea
\label{eq:Tdual}
& & H_{ijk} \longrightarrow \omega_{ij}{}^k  \longrightarrow Q^{jk}{}_i  \longrightarrow R^{ijk} \, .
\eea
Such fluxes are interpreted as possible gaugings in the gauges supergravity picture \cite{Derendinger:2004jn,Grana:2012rr,Dibitetto:2012rk,Aldazabal:2006up}. Turning on these fluxes on compactification background, the 4D-effective potential generically depends on all of  such fluxes as a set of parameters, and have been useful for studying moduli stabilization and string vacua \cite{Derendinger:2004jn,Grana:2012rr,Dibitetto:2012rk,Kachru:2002sk, Hellerman:2002ax,Dabholkar:2002sy, Danielsson:2012by, Blaback:2013ht, Damian:2013dq, Damian:2013dwa, Hassler:2014mla, Ihl:2007ah, deCarlos:2009qm, Blumenhagen:2015qda, Blumenhagen:2015kja}. 

It is surprisingly remarkable that the non-geometric 4D effective potentials could be studied via merely knowing the forms of K\"ahler and super-potentials \cite{Danielsson:2012by, Villadoro:2005cu, Blumenhagen:2013hva, Gao:2015nra, Shukla:2015bca, Ihl:2007ah, deCarlos:2009qm, Robbins:2007yv}, and without having a full understanding of their ten-dimensional origin. Some significant steps have been taken towards exploring the form of non-geometric 10D action via Double Field Theory (DFT) \cite{Andriot:2013xca, Andriot:2011uh, Andriot:2012wx} as well as supergravity \cite{Villadoro:2005cu, Blumenhagen:2013hva, Gao:2015nra}. In this regard, a ten-dimensional origin of the four dimensional  scalar potential with geometric flux $(\omega)$ in type IIA toroidal orietifold has been invoked in \cite{Villadoro:2005cu}. This has been subsequently generalized with the inclusion of non-geometric $(Q, R)$-fluxes for type IIA- and non-geometric $(Q)$-flux for IIB-theory in \cite{Blumenhagen:2013hva} and the resulting oxidized 10D action was found to be compatible with  DFT action. %\cite{Aldazabal:2011nj,Geissbuhler:2011mx}.

The interest towards studying non-geometric flux has been extended while ensuring the modular invariance of the underlying type IIB supergravity, and the need of additional non-geometric flux has been argued \cite{Aldazabal:2008zza,Font:2008vd, Guarino:2008ik}. This has led to compactification manifold to be a $U$-fold in which the local patched are glued via performing $T$- and $S$-dualities \cite{Hull:2004in, Aldazabal:2008zza, Font:2008vd, Kumar:1996zx, Hull:2003kr, Chatzistavrakidis:2013jqa}. Also, the flux-orbits of \cite{Blumenhagen:2013hva} have been further generalized with non-geometric $P$-flux dual to $Q$-flux \cite{Gao:2015nra}. Moreover, it has been observed \cite{Blumenhagen:2013hva, Gao:2015nra}  that various NS-NS/RR (non-)geometric fluxes,  appear with generalized flux combinations. 

In the conventional approach of studying 4D type II effective theories in a non-geometric flux compactification framwork, most of the studies have been centered around ${\mathbb T}^6/({\mathbb Z}_2 \times {\mathbb Z}_2)$ orientifold. Though very useful for learning many aspects, setups based on this orientifold are too simple to reflect all the features; for example the absence of odd axions (at least with the known involutions). Given that axionic inflationary models have attracted a lot of attention in recent times, and generic string compactifications naturally result in many axions arising from NS-NS and RR form-fields in ten dimensions, it is desired to extend the previous studies on non-geometric flux compactification (and its moduli stabilization applicabilities into less simple toroidal or CY orientifold setups) to include all possible axions to play with. Such an initiative has been taken in \cite{Robbins:2007yv} by considering a  ${\mathbb T}^6/{\mathbb Z}_4$ toroidal orientifold\footnote{For explicit construction of more type IIB toroidal/CY orientifold setups with odd-axions, see \cite{Lust:2006zg,Lust:2006zh,Blumenhagen:2008zz,Cicoli:2012vw,Cicoli:2012bi,Gao:2013rra,Gao:2013pra}.} which has been further utilized for studying the dimensional oxidation of four dimensional non-geometric effective potential in \cite{Shukla:2015bca}. {\it In this article, we provide a modular completed version of a new set of generalized flux-orbits which include all the (non-)geometric fluxes along with the odd $B_2/C_2$-axions.} We will exemplify the utility of those new generalized orbits in a couple of concrete examples.

The article is organized as follows: In section \ref{sec_setup}, we recollect the relevant preliminaries of a type IIB non-geometric flux compactification setup. In section \ref{sec_invokingOrbits}, we will present an intuitive search for the new generalized flux combinations via closely looking at the superpotential and D-term contributions. Subsequently, in section \ref{sec_rearrangement}, we will exemplify the utility of these modular completed new flux orbits in rewriting the four dimensional effective scalar potential in a very compact manner. Finally, in section \ref{sec_conclusions} we summarize the results with outlook.    

\section{Non-geometric flux compactification and type IIB orientifolds}
\label{sec_setup}
\subsection{Fixing the conventions}
We consider Type IIB superstring theory compactified on an orientifold of a
Calabi-Yau threefold $X$.
The admissible orientifold projections can be classified by their action on the
K\"ahler form $J$ and the holomorphic three-form $\Omega_3$ of
the Calabi-Yau, given as under \cite{Grimm:2004uq}:
\begin{eqnarray}
\label{eq:orientifold}
 {\cal O}= \begin{cases}
                       \Omega_p\, \sigma  & : \, 
                       \sigma^*(J)=J\,,  \, \, \sigma^*(\Omega_3)=\Omega_3 \, ,\\[0.1cm]
                       (-)^{F_L}\,\Omega_p\, \sigma & :\, 
        \sigma^*(J)=J\,, \, \, \sigma^*(\Omega_3)=-\Omega_3\, ,
\end{cases}
\end{eqnarray}
where $\Omega_p$ is the world-sheet parity, $F_L$ is the left-moving space-time fermion number, and $\sigma$ is a holomorphic, isometric
 involution. The first choice leads to orientifold with $O5/O9$-planes
whereas the second choice to $O3/O7$-planes.
%The generated R-R tadpoles need to be cancelled by the introduction 
%of the corresponding D-branes. For latter case, the one of primary
%interest here,  these are in general
%$D7$-branes carrying addition gauge flux and $D3$-branes.
%The massless bosonic spectrum of the ten-dimensional type IIB theory includes
%the dilaton $\phi$ with an axion $C_0$, the metric $g$, an NS-NS two-form
%$B_2$, RR forms $C_2$ and $C_4$, which has a self-dual field strength in the
%R-R sector.
%The $(-)^{F_L}\,\Omega_p\, \sigma$  invariant states in four-dimensions
%are given as: 
%in table \ref{tableprojection}.
%\begin{table}[H]
%  \centering
%  \begin{tabular}{c||c|c|c|c|c|c}
%  & $\phi$&  $g_{\mu \nu}$ &  $B_2$&  $C_0$ & $C_2$ & $C_4$\\
%  \hline \hline
% $(-)^{F_L}$ &  $+$  &   $+$  &  $+$   &   $-$    &   $-$  & $-$\\
% $\Omega_p$ &   $+$  &   $+$  &  $-$   &   $-$    &   $+$  & $-$ \\
% $\sigma^*$ &   $+$  &   $+$  &  $-$   &   $+$    &   $-$  & $+$ \\
% \end{tabular}
%  \caption{\small Orientifold invariant states.}
%  \label{tableprojection}
%\end{table}
{\vskip-0.0cm The massless states in the four dimensional effective theory are in one-to-one correspondence
with harmonic forms which are either  even or odd
under the action of $\sigma$, and these do generate the equivariant  cohomology groups $H^{p,q}_\pm (X)$. Let us fix our conventions as those of \cite{Robbins:2007yv}, and denote the bases  of even/odd two-forms as $(\mu_\alpha, \, \nu_a)$ while four-forms as $(\tilde{\mu}_\alpha, \, \tilde{\nu}_a)$ where $\alpha\in h^{1,1}_+(X), \, a\in h^{1,1}_-(X)$. Also, we denote the zero- and six- even forms as ${\bf 1}$ and $\Phi_6$ respectively. The definitions of integration over the intersection of various cohomology bases are,
\bea
\label{eq:intersection}
& & \hskip-0.7cm \int_X \Phi_6 = f, \, \, \int_X \, \mu_\alpha \wedge \tilde{\mu}^\beta = \hat{d}_\alpha^{\, \, \, \beta} , \, \, \int_X \, \nu_a \wedge \tilde{\nu}^b = {d}_a^{\, \, \,b} \\
& & \hskip-0.7cm \int_X \, \mu_\alpha \wedge \mu_\beta \wedge \mu_\gamma = k_{\alpha \beta \gamma}, \, \, \, \, \int_X \, \mu_\alpha \wedge \nu_a \wedge \nu_b = \hat{k}_{\alpha a b} \nonumber
\eea
%In addition to the splitting of $H^2(X)$ and its dual $H^4(X)$-cohomologies, we also need to know the splitting of three-form cohomology $H^3(X)$ into even/odd eigenspaces under a given involution $\sigma$. 
Note that if four-form basis is chosen to be dual of the two-form basis, one will of course have $\hat{d}_\alpha^{\, \, \, \beta} = \hat{\delta}_\alpha^{\, \, \, \beta}$ and ${d}_a^{\, \, \,b} = {\delta}_a^{\, \, \,b}$. However for the present work, we follow the conventions of \cite{Robbins:2007yv}, and take a bit more generic case. Considering the bases for the even/odd cohomologies $H^3_\pm(X)$ of three-forms as symplectic pairs $(a_K, b^K)$ and $({\cal A}_k, {\cal B}^k)$ respectively, we fix the normalization as under,
\bea
\int_X a_K \wedge b^J = \delta_K{}^J, \, \, \, \, \, \int_X {\cal A}_k \wedge {\cal B}^j = \delta_k{}^j
\eea
Here, for the orientifold choice with $O3/O7$-planes, $K\in \{1, ..., h^{2,1}_+\}$ and $k\in \{0, ..., h^{2,1}_-\}$ while for $O5/O9$-planes, one has $K\in \{0, ..., h^{2,1}_+\}$ and $k\in \{1, ..., h^{2,1}_-\}$.

Now, the various field ingredients can be expanded in appropriate bases of the equivariant cohomologies. For example, the K\"{a}hler form $J$, the
two-forms $B_2$,  $C_2$ and the R-R four-form $C_4$ can be expanded as \cite{Grimm:2004uq}
\bea
\label{eq:fieldExpansions}
& & J = t^\alpha\, \mu_\alpha,  \,\,\,\quad  B_2= b^a\, \nu_a , \,\,\, \quad C_2 =c^a\, \nu_a \, \\
& & C_4 = D_2^{\alpha}\wedge \mu_\alpha + V^{K}\wedge a_K + U_{K}\wedge b^K + {\rho}_{\alpha} \, \tilde\mu^\alpha \nonumber
\eea
where $t^\alpha$ is string-frame two-cycle volume moduli, while $b^a, \, c^a$ and $\rho_\alpha$ are various axions. Further, ($V^K$, $U_K$) forms a dual pair of space-time one-forms
and $D_2^{\alpha}$ is a space-time two-form dual to the scalar field $\rho_\alpha$.
%Due to the self-duality of RR four-form, half of the degrees of freedom of $C_4$ are removed.
%Note that the even component of the Kalb-Ramond field $B_{+} = b^\alpha \, \omega_\alpha$, though  not a continuous modulus, can take the two  discrete values $b^\alpha\in\{0, 1/2\}$. 
Further, since $\sigma^*$ reflects the holomorphic three-form $\Omega_3$, we have $h^{2,1}_-(X)$ complex structure moduli $z^{\tilde a}$ appearing as complex scalars. %, and finally one has the following table summarizing the ${\cal N}=1$ supersymmetric massless bosonic spectrum \cite{Grimm:2004uq}, 
%\begin{table}[H]
%\centering
%\begin{tabular}{|l|l|l|}
%\hline
% &  & \\[-0.2cm]
%  & $h_-^{2,1}$ & $z^{\tilde a}$ \\
%chiral multiplets & $h_+^{1,1}$ & $(t^\alpha, \rho_{\alpha})$ \\
%  & $h_-^{1,1}$ & $(b^a, c^a)$\\
%   & 1 & $(\phi, C_0)$\\
%    &  & \\[-0.2cm]
%    \hline
%    &  & \\[-0.2cm]
%vector multiplet & $h_+^{2,1}$ & $V^K$ \\
%&  & \\[-0.2cm]
%\hline
% &  & \\[-0.2cm]
%gravity multiplet & 1 & $g_{\mu\nu}$ \\[0,2cm]
%\hline
%\end{tabular}
%\caption{${\cal N}=1$ massless bosonic spectrum of Type IIB Calabi Yau orientifold
%\label{table_susy}}
%\end{table}
%\noindent
Now, we consider a complex multi-form of even degree $\Phi_c^{even}$ defined as \cite{Benmachiche:2006df},
\bea
& & \hskip-1cm \Phi_c^{even} = e^{B_2} \wedge C_{RR} + i \, e^{-\phi} Re(e^{B_2+i\, J})\\
%& & \hskip-1cm  =(C_0 + i \, e^{-\phi}) + \left(C_2 + (C_0 + i \, e^{-\phi}) B_2\right) \nonumber\\
%& & + \left(C_4^{(0)}+C_2\wedge B_2 + \frac{1}{2} (C_0 + i \, e^{-\phi}) B_2 \wedge B_2 - \frac{i}{2} e^{-\phi} J \wedge J\right)\nonumber \\
& & \hskip-0.0cm \equiv \tau + G^a \, \,\nu_a + T_\alpha \, \, \tilde{\mu}^\alpha \, ,\nonumber
\eea
which suggests the following forms for the Einstein-frame chiral variables appearing in ${\cal N}=1$ 4D-effective theory,
\bea
\label{eq:N=1_coords}
& & \tau = C_0 + \, i \, e^{-\phi}\, \, , \, \, \, \, \, G^a:= c^a + \tau \, b^a \, ,\\
& & \hskip-0.05cm T_\alpha:= \left({\rho}_\alpha +  \hat{\kappa}_{\alpha a b} c^a b^b + \frac{1}{2} \, \tau \, \hat{\kappa}_{\alpha a b} b^a \, b^b \right)  -\frac{i}{2} \, \kappa_{\alpha\beta\gamma} t^\beta t^\gamma\, ,\nonumber
\eea
where $\kappa_{\alpha\beta\gamma}=(\hat{d^{-1}})_\alpha^{ \, \,\delta} \, k_{\delta\beta\gamma}$ and $\hat{\kappa}_{\alpha a b} = (\hat{d^{-1}})_\alpha^{ \, \,\delta} \, \hat{k}_{\delta a b}$. From now onwards we will use $e^{-\phi} = s$.
\subsection{Four dimensional effective theory}
The dynamics of low energy effective supergravity action is encoded in three building blocks; namely a K\"{a}hler potential $K$, a holomorphic superpotential $W$ and a
holomorphic gauge kinetic function $\hat{\cal G}$ written in terms of the appropriate chiral variables defined in eqn. (\ref{eq:N=1_coords}). The four-dimensional scalar potential receives contributions from F-terms as well as D-terms and can be denoted as,
\bea
& & V= V_F + V_D 
\eea

\subsubsection*{F-term contributions}
The F-term contributions to the ${\cal N}=1$ scalar potential are computed from the K\"ahler and super-potential via
\bea
\label{eq:Vtot}
& & V_F=e^{K}\Big(K^{i\bar\jmath}D_i W\, D_{\bar\jmath} \ov W-3\, |W|^2\Big) \nonumber
\eea
Using appropriate chiral variables, a generic form of K\"{a}hler potential (at tree level) can be given as,
\bea
\label{eq:K}
& & \hskip-0.49cm K = - \ln\left(-i(\tau-\ov\tau)\right)-\ln\left(i\int_{X}\Omega_3\wedge{\bar\Omega_3}\right) -2\ln{\cal V}_E,
\eea
%\bea
%\label{eq:K}
%& & \hskip-0.5cm K = - \ln\left(-i(\tau-\ov\tau)\right)-\ln\left(i\int_{X}\Omega_3\wedge{\bar\Omega_3}\right)\\
%& & -2\ln\left({\cal V}_E\,(\tau, G^a, T_\alpha; \ov \tau, \ov G^a, \ov T_\alpha)\right) \nonumber
%\eea
where ${\cal V}_E$ is the Einstein frame volume of the CY.%, and $\Omega_3$ denotes the involutively-odd nowhere vanishing holomorphic 3-form. 

To understand the splitting of various geometric as well as non-geometric fluxes into the suitable orientifold even/odd bases, it is important to note that in a given setup, all flux-components will not be generically allowed under the full orietifold action ${\cal O} = \Omega_p (-)^{F_L} \sigma$. For example, only geometric flux $\omega$ and non-geometric flux $R$ remain invariant under  $(\Omega_p (-)^{F_L})$, while the standard fluxes $(F, H)$ and non-geometric flux $(Q)$ are anti-invariant \cite{Blumenhagen:2015kja, Robbins:2007yv}. Therefore, under the full orientifold action, we can only have the following flux-components %of the standard fluxes $(F, H)$ and the geometric $(\omega)$ as well as non-geometric fluxes $(Q$ and $R$),
\bea
& & \hskip-0.10cm F\equiv \left(F_k, F^k\right),  H\equiv \left(H_k, H^k\right), \omega\equiv \left({\omega}_a{}^k, {\omega}_{a k} , \hat{\omega}_\alpha{}^K, \hat{\omega}_{\alpha K}\right),\nonumber\\
& &  R\equiv \left(R_K, R^K \right), \, \, \, Q\equiv \left({Q}^{a{}K}, \, {Q}^{a}{}_{K}, \, \hat{Q}^{\alpha{}k} , \, \hat{Q}^{\alpha}{}_{k}\right), 
%& &  P\equiv \left({P}^{a{}K}, \, {P}^{a}{}_{K}, \, \hat{P}^{\alpha{}k} , \, \hat{P}^{\alpha}{}_{k}\right)
\eea
For writing a general flux-superpotential, one needs to define a twisted differential operator, ${\cal D}$ involving the actions from all the NS-NS (non-)geometric fluxes as \cite{Robbins:2007yv}, 
\bea
\label{eq:twistedD}
& & {\cal D} = d + H \wedge.  - \omega \triangleleft . + Q \triangleright. - R \bullet \, \, .
\eea
The action of operator $\triangleleft, \triangleright$ and $\bullet$ on a $p$-form changes it into a $(p+1)$, $(p-1)$ and $(p-3)$-form respectively. Considering various flux-actions on the different  even/odd bases to result in even/odd three-forms, we have \cite{Robbins:2007yv},
%\begin{subequations}
\bea
\label{eq:action1}
& & H = {H}^k {\cal A}_k + H_k \, \,{\cal B}^k, \, \, \, \, \, F = {F}^k {\cal A}_k + F_k \, \,{\cal B}^k,\nonumber\\
& & \omega_a \equiv (\omega \triangleleft \nu_a) = {\omega}_a{}^k \, {\cal A}_k + \omega_{a{}k} {\cal B}^k, \nonumber\\
& & \hat{Q}^{\alpha}\equiv (Q \triangleright {\tilde\mu}^\alpha) = \hat{Q}^{\alpha{}k} {\cal A}_k + \hat{Q}^{\alpha}{}_{k} {\cal B}^k \\
%& & \hat{P}^{\alpha}\equiv (P \triangleright {\tilde\mu}^\alpha) = \hat{P}^{\alpha{}k} {\cal A}_k + \hat{P}^{\alpha}{}_{k} {\cal B}^k \nonumber
%\eea
%and
%\bea
& & \hat{\omega}_\alpha \equiv (\omega \triangleleft \mu_\alpha) = \hat{\omega}_\alpha{}^K a_K + \hat{\omega}_{\alpha{}K} b^K, \nonumber\\
& & {Q}^{a}\equiv (Q \triangleright \tilde{\nu}^a) = {Q}^{a{}K} \, a_K + Q^{a}{}_{K} b^K, \nonumber\\
%& & {P}^{a}\equiv (Q \triangleright \tilde{\nu}^a) = {P}^{a{}K} \, a_K + P^{a}{}_{K} b^K, \nonumber\\
& & R\bullet \Phi = R^K a_K + R_K b^K \, . \nonumber
\eea
%\end{subequations}
The first three lines involve flux components counted via `odd-index' $k\in{h^{2,1}_-(X)}$ while the later three have `even-index' $K\in {h^{2,1}_+(X)}$. Using definitions in (\ref{eq:action1}), we have the following additional useful non-trivial actions of fluxes on various 3-form even/odd basis elements \cite{Robbins:2007yv},
\bea
\label{eq:action2}
& & \hskip-0.04cm H \wedge {\cal A}_k = - f^{-1} H_k \Phi_6, \, \, \quad  \, \, H \wedge {\cal B}^k = f^{-1} H^k \Phi_6 \\
& & \hskip-0.04cm \omega\triangleleft {\cal A}_k=\left({d}^{-1}\right)_a{}^b \,{\omega}_{b k} \, \tilde{\nu}^a, \, \quad   \omega\triangleleft {\cal B}^k=-\left({d}^{-1}\right)_a{}^b \, {\omega}_{b}{}^{k} \, \tilde{\nu}^a\nonumber\\
& & \hskip-0.04cm Q\triangleright {\cal A}_k=-\left(\hat{d}^{-1}\right)_\alpha{}^\beta \,\hat{Q}^\alpha_{k} \, {\mu}_\beta, \, \, Q\triangleright {\cal B}^k=\left(\hat{d}^{-1}\right)_\alpha{}^\beta \, \hat{Q}^{\alpha k} \,{\mu}_\beta ,\nonumber\\
%& & \hskip-0.4cm P\triangleright {\cal A}_k=-\left(\hat{d}^{-1}\right)_\alpha{}^\beta \,\hat{P}^\alpha_{k} \, {\mu}_\beta, \, \, P\triangleright {\cal B}^k=\left(\hat{d}^{-1}\right)_\alpha{}^\beta \, \hat{P}^{\alpha k} \,{\mu}_\beta ,\nonumber
%\eea
%and
%\bea
%\label{eq:action3}
& & \hskip-0.04cm R\bullet a_K = f^{-1} \, R_K \, {\bf 1}, \, \, \,\quad  \, \, R \bullet b^K = - f^{-1} \, R^K \,{\bf 1}\nonumber\\
& & \hskip-0.04cm \omega\triangleleft a_K=\left(\hat{d}^{-1}\right)_\alpha{}^\beta \hat{\omega}_{\beta K} \, \tilde{\mu}^\alpha, \,  \, \omega\triangleleft b^K=-\left(\hat{d}^{-1}\right)_\alpha{}^\beta \hat{\omega}_{\beta}{}^{K} \, \tilde{\mu}^\alpha\nonumber\\
& & \hskip-0.04cm Q\triangleright a_K=-\left({d}^{-1}\right)_a{}^b {Q}^a_{K} \,{\nu}_b, \,\, Q\triangleright b^K=\left({d}^{-1}\right)_a{}^b \,{Q}^{a K} \, {\nu}_b \, .\nonumber
%& & \hskip-0.4cm P\triangleright a_K=-\left({d}^{-1}\right)_a{}^b {P}^a_{K} \,{\nu}_b, \,\, P\triangleright b^K=\left({d}^{-1}\right)_a{}^b \,{P}^{a K} \, {\nu}_b . \nonumber
\eea
With these ingredients in hand, a generic form of flux superpotential is as under,
\bea
\label{eq:W1}
& & \hskip-0.5cm W =- \int_X \biggl[F+ {\cal D} \Phi_c^{even}\biggr]_3 \wedge \Omega_3  = - \int_{X} \biggl[{F} +\tau \, {H} + \, \omega_a {G}^a + \, {\hat Q}^{\alpha} \,{T}_\alpha \biggr]_3 \wedge \Omega_3. 
\eea
Note that, only $\omega_a$ and $\hat{Q}^\alpha$ components are allowed by the choice of involution to contribute into the superpotential, and in order to turn-on the non-geometric $R$-fluxes, one has to induce D-terms via implementing a non-trivial even sector of $H^{2,1}(X)$-cohomology as we will explain now.
\subsubsection*{D-term contributions}
Apart from the usual brane/orientifold local source contributions needed to cancel tadpoles, in the presence of non-trivial even sector of $H^{2,1}(X)$-cohomology, generically there are additional D-terms possible to contribute (say ${V_D}^{(1)}$) to the four dimensional scalar potential given as \cite{Robbins:2007yv}, 
\bea
& & \hskip-0.75cm {V_D}^{(1)} = \frac{1}{2} (Re \, \, \, {\mathbb F})_e^{{-1}{JK}} \, D_J D_K + \frac{1}{2} (Re \, \, \, {\mathbb F})_m^{-1}{}_{{JK}} \, D^J D^K,
\eea
where $(Re \, {\mathbb F})_{e/m}^{{-1}{JK}}$ represent the electric/magnetic gauge-kinetic couplings. For D-term components $(D_K, D^K)$, one can follow the strategy of \cite{Robbins:2007yv} via considering the gauge transformations of RR potentials $C_{RR} \equiv \left(C_0 + c^a \nu_a + \rho_\alpha \tilde{\mu}^\alpha \right) \to  C_{RR} + {\cal D} (\lambda^K a_K + \lambda_K b^K )$ to get,
%let us consider the following gauge transformations of RR potentials $C_{RR} = C_0 + C_2 + C_4$,
%\bea
%& & \hskip-0.5cm C_{RR} \equiv \left(C_0 + c^a \nu_a + \rho_\alpha \tilde{\mu}^\alpha \right) \to  C_{RR} + {\cal D} (\lambda^K a_K + \lambda_K b^K )\nonumber\\
%& & \hskip-0.5cm = \left(C_0 - f^{-1} R_K \lambda^K + f^{-1} R^K \lambda_K\right) + \biggl(c^b -(d^{-1})_a{}^b Q^a{}_K \lambda^K \nonumber\\
%& & + (d^{-1})_a{}^b Q^{a K} \lambda_K \biggr) \nu_b + \biggl(\rho_\alpha -({\hat{d}}^{-1})_\alpha{}^\beta \, \hat{\omega}_{\beta K} \lambda^K \\
%& & + ({\hat{d}}^{-1})_\alpha{}^\beta \, \hat{\omega}_{\beta}{}^{K} \lambda_K\biggr) \tilde{\mu}^\alpha\nonumber
%\eea
%This suggests the following two D-terms being generated by the gauge transformations,
\bea
\label{eq:D-terms}
& & \hskip-0.8cm D_K = -i \, \biggl[ f^{-1} R_K \, (\partial_\tau K) + (d^{-1})_b{}^a Q^b{}_K \, (\partial_a K) + ({\hat{d}}^{-1})_\alpha{}^\beta \, \hat{\omega}_{\beta K}\, (\partial^\alpha K) \biggr], \\
& & \hskip-0.8cm D^K = i \, \biggl[ f^{-1} R^K \, (\partial_\tau K) + (d^{-1})_b{}^a Q^{b K} \, (\partial_a K)  + ({\hat{d}}^{-1})_\alpha{}^\beta \, \hat{\omega}_{\beta}{}^{K}\, (\partial^\alpha K) \biggr] \nonumber
\eea
Now although it is not possible to explicitly write down the overall internal volume ${\cal V}_E$ in terms of chiral variables $(T_\alpha, G^a, \tau)$ for a generic CY orientifold case, however one can still compute the K\"ahler derivatives as well as (inverse-)metric \cite{Grimm:2004uq}. Using the expressions for the generic tree level K\"ahler potential (\ref{eq:K}), one finds,  
\bea
& & \hskip-1.2cm \partial_\tau K = \frac{i}{2\, s\, {\cal V}_E} \left({\cal V}_E - \frac{s}{2}\hat{k}_{\alpha a b} t^\alpha b^a b^b \right), \quad \partial_{G^a} K = \frac{i}{2 \, {\cal V}_E}\hat{k}_{\alpha a b} t^\alpha b^b, \quad \partial_{T_\alpha} K = -\frac{i \, {\hat{d}}^\alpha{}_\beta \, t^\beta}{2 \, {\cal V}_E} 
\eea
Subsequently, we have
\bea
\label{eq:DtermOld}
& & \hskip-1.2cm \quad D_K = \frac{1}{2\, s\,{\cal V}_E}\, \biggl[ \frac{R_K}{f} \, \left({\cal V}_E -\frac{s}{2}\hat{k}_{\alpha a b} t^\alpha b^a b^b\right) + s\, (d^{-1})_b{}^a Q^b{}_K \, \hat{k}_{\alpha a c} t^\alpha b^c - s\, t^\alpha \, \,\hat{\omega}_{\alpha K}\, \biggr]\\
& & \hskip-1.2cm \quad D^K = -\frac{1}{2\, s\,{\cal V}_E}\, \biggl[ \frac{R^K}{f} \, \left({\cal V}_E -\frac{s}{2}\hat{k}_{\alpha a b} t^\alpha b^a b^b\right) + s\, (d^{-1})_b{}^a Q^{b K} \, \hat{k}_{\alpha a c} t^\alpha b^c - s\, t^\alpha \, \,\hat{\omega}_{\alpha}{}^{K}\, \biggr]\nonumber
\eea
Further, taking into account the contributions (say ${V_D^{(2)}}$) coming from local sources such as branes and orientifolds to cancel the RR tadpoles, the total four dimensional scalar potential can be given as,
\bea
& & V_{\rm tot} = V_F + {V_D}^{(1)} + {V_D}^{(2)}.
\eea
%Note that, in conventional phenomenological approach, e.g. via moduli stabilization in LVS \cite{} or KKLT scenarios \cite{}, one considers the orientifold involution such that $h^{2,1}(CY/{\cal O}) = h^{2,1}_-(CY/{\cal O}) $, i.e. $h^{2,1}_+(CY/{\cal O})= 0$, and so in those approaches, no $D$-terms of type ${V_D}^{(1)}$ are induced, and so the non-geometric $R$-fluxes will be absent.
\subsection{Modular completion}
The four dimensional scalar potential generically have an S-duality invariance following from the underlying ten-dimensional type IIB supergravity. The same corresponds to the following $SL(2, \mathbb{Z})$ transformation,
\bea
\label{eq:SL2Za}
& & \hskip-1.5cm \tau\to \frac{a \tau+ b}{c \tau + d}\, \quad \quad {\rm where} \quad a d- b c = 1\,;\quad a,\ b,\ c,\ d\in \mathbb{Z}
\eea
Under this $SL(2, \mathbb{Z})$ transformation, noting that complex structure moduli and the Einstein frame Calabi Yau volume (${\cal V}_E$) are invariant, the K\"ahler potential (at the tree level) given in eqn. (\ref{eq:K}) transform as:
\bea
\label{eq:S-duality1}
e^K \longrightarrow |c \, \tau + d|^2 \, e^K\, , 
\eea
and therefore the S-duality invariance of physical quantities (e.g. gravitino mass-square which involves a factor of $e^K |W|^2$-type) suggests that the holomorphic superpotential, $W$ should have a modularity of weight $-1$, and so we have \cite{Font:1990gx, Cvetic:1991qm, Grimm:2007xm}
\bea
\label{eq:modularW}
& & W \to \frac{W}{c \, \tau + d}
\eea
Now, in a given flux compactification scenario, the various fluxes have to readjust among themselves to respect this modularity condition (\ref{eq:modularW}). For example, in the absence of geometric flux ($\omega$) and the non-geometric flux ($Q$), the three-form combination $(F_3 + \tau H_3)$ appearing in the superpotential (\ref{eq:W1}) can ensure the condition (\ref{eq:modularW}) to hold if the standard NS-NS and R-R fluxes, $H_3$ and $F_3$ adjust themselves as under,
\bea
\label{eq:SL2Zfa}
& &  \begin{pmatrix}F_3\\ H_3\end{pmatrix}\to\begin{pmatrix}a&-b\\
-c&d\end{pmatrix}\begin{pmatrix}F_3\\ H_3\end{pmatrix}.
\eea
Recall that we are following the conventions of \cite{Robbins:2007yv} which involves a sign flip in $B_2$ (and hence in $H_3$) as compared to those of \cite{Benmachiche:2006df,Grimm:2007xm}. Therefore, a minus sign with $b$ and $c$ appears in the above transformation of $H_3$ and $F_3$ fluxes. Nevertheless it does not matter as the $SL(2, \IZ)$ condition $ad - bc = 1$ is unaffected.

Moreover, the  $SL(2, \IZ)$ self-dual action in (\ref{eq:SL2Za}) will result in further transformation on the rest of the massless bosonic spectrum. The chiral variables (other than axion-dilaton) transform as under \cite{Grimm:2007xm},
\bea
\label{eq:SL2Zb}
& & G^a \to \frac{G^a}{c \tau + d}, \quad \quad T_\alpha \to T_\alpha - \frac{c}{c\tau + d} \left(\frac{1}{2}\hat{\kappa}_{\alpha a b} \, G^a \, G^b \right),
\eea
where in order to verify the above transformations, one would need the following,
\bea
\label{eq:SL2Zc}
& & \hskip-1.5cm   \begin{pmatrix}C_2\\ B_2\end{pmatrix}\to\begin{pmatrix}a&-b\\
-c&d\end{pmatrix}\begin{pmatrix}C_2\\ B_2\end{pmatrix}\,
\eea
On the similar lines as to those of $(F_3, H_3)$ doublet, the inclusion of P-flux, which is S-dual to the standard non-geometric $Q$-flux, provides a modular completion under the $SL(2, \IZ)$ transformation \cite{Aldazabal:2006up}:
\bea
\label{eq:SL2Zfb}
& & \begin{pmatrix}Q\\ P\end{pmatrix}\to\begin{pmatrix}a&-b\\
-c&d\end{pmatrix}\begin{pmatrix}Q\\ P\end{pmatrix}.
\eea
The $SL(2, \IZ)$ transformations $\begin{pmatrix}a&-b\\
-c&d\end{pmatrix}$ involved in the flux adjustments have two generators $g_1$ and $g_2$ corresponding to two physically different cases,
\begin{itemize}
\item{First transformation:
\bea
& & \hskip-1cm ~\tau \to -\frac{1}{\tau} \Longleftrightarrow \{a=0=d, b \, c = -1\}, \quad \quad \quad g_1 = \begin{pmatrix}0&-1\\
1&0\end{pmatrix}\, 
\eea
Subsequently, one finds that
\bea
\label{eq:S-duality}
& &  \hskip-1.5cm B_2 \to C_2, \quad C_2 \to - \, B_2, \quad \quad \quad G^a \to - \, \frac{G^a}{\tau},  \quad \quad \quad T_{\alpha} \to {T_{\alpha}}- \frac{1}{2} \frac{\hat{\kappa}_{\alpha a b} G^a G^b}{\tau}\\
%& & C_4 \to C_4, \,\,C_8 \to \tilde C_8,\,\, \tilde C_8 \to C_8,\,\, C'_8 \to - C'_8,\\
& & \hskip-1.0cm {H}_{ijk} \to {F}_{ijk}, \quad {F}_{ijk} \to -{H}_{ijk}, \quad \quad  \quad {Q}^{ij}_{k} \to - {P}^{ij}_{k}, \quad {P}^{ij}_{k}\to {Q}^{ij}_{k} \,. \nonumber
\eea}This first transformation corresponds to the popularly known strong-weak duality while the second one given as under corresponds to a shift in the universal axion $C_0 \to C_0 + 1$. 
\item{Second transformation\footnote{We thank the referee for suggesting us to look at the effect of $\tau \to \tau +1$ transformation.}:
\bea
& & \hskip-1cm ~\tau \to \tau + 1 \Longleftrightarrow \{c=0, a=b=d, a\, d = 1\}, \quad \quad \quad g_2 = \begin{pmatrix}1&-1\\
0&1\end{pmatrix}
\eea
Subsequently, one finds that
\bea
\label{eq:S-duality1}
& &  \hskip-1.5cm B_2 \to B_2, \quad C_2 \to C_2 - \, B_2, \quad \quad \quad G^a \to \, {G^a},  \quad \quad \quad T_{\alpha} \to {T_{\alpha}}\\
%& & C_4 \to C_4, \,\,C_8 \to \tilde C_8,\,\, \tilde C_8 \to C_8,\,\, C'_8 \to - C'_8,\\
& & \hskip-2.5cm {H}_{ijk} \to {H}_{ijk}, \quad {F}_{ijk} \to {F}_{ijk}-{H}_{ijk}, \quad \quad  \quad {Q}^{ij}_{k} \to {Q}^{ij}_{k} - {P}^{ij}_{k}, \quad {P}^{ij}_{k}\to {P}^{ij}_{k} \,. \nonumber
\eea}
\end{itemize}
From the point of view of K\"ahler potential and the superpotential, the later case amounts to just some rescalings $e^K \to |d|^2\, e^K$ and $W \to W/d$ as $\tau \to \tau +1$ simply implies $c=0$. Now similar to Q-flux actions given in eqns. (\ref{eq:action1})-(\ref{eq:action2}), the orientifold invariance allows the P-fluxes of only $({P}^{a{}K}, \, {P}^{a}{}_{K}, \, \hat{P}^{\alpha{}k} , \, \hat{P}^{\alpha}{}_{k})$ types with their actions being, 
%\begin{subequations}
\bea
& & \hskip1.75cm \hat{P}^{\alpha}\equiv (P \triangleright {\tilde\mu}^\alpha) = \hat{P}^{\alpha{}k} {\cal A}_k + \hat{P}^{\alpha}{}_{k} {\cal B}^k\, , \\
& & \hskip1.75cm {P}^{a}\equiv (Q \triangleright \tilde{\nu}^a) = {P}^{a{}K} \, a_K + P^{a}{}_{K} b^K\, , \nonumber\\
& & \hskip-0.5cm P\triangleright {\cal A}_k=-\left(\hat{d}^{-1}\right)_\alpha{}^\beta \,\hat{P}^\alpha_{k} \, {\mu}_\beta, \quad \quad P\triangleright {\cal B}^k=\left(\hat{d}^{-1}\right)_\alpha{}^\beta \, \hat{P}^{\alpha k} \,{\mu}_\beta \, , \nonumber\\
& & \hskip-0.5cm P\triangleright a_K=-\left({d}^{-1}\right)_a{}^b {P}^a_{K} \,{\nu}_b, \quad \quad \quad P\triangleright b^K=\left({d}^{-1}\right)_a{}^b \,{P}^{a K} \, {\nu}_b . \nonumber
\eea
%\end{subequations}
Subsequently, a modular completed form of superpotential (\ref{eq:W1}) is given as below \cite{Blumenhagen:2015kja},
\bea
\label{eq:W2}
& & \hskip-0.99cm W = - \int_{X} \biggl[\left({F} +\tau \, {H}\right) + \, \omega_a  \, {G}^a + \, \left(\hat{Q}^{\alpha}+ \tau\, \hat{P}^{\alpha}\right) \,{T}_\alpha  \nonumber\\
& & \hskip1.00cm -\,\hat{P}^{\alpha} \left(\frac{1}{2} \hat{\kappa}_{\alpha a b} G^a G^b\right) \biggr]_3 \wedge \Omega_3. 
\eea
Note that, $R$-flux does not appear in the superpotential \cite{Robbins:2007yv, Blumenhagen:2015kja}, and hence in the $F$-term scalar potential as well. In addition, we observe that for the superpotential in eqn. (\ref{eq:W2}), one finds that $W \to -\frac{W}{\tau}$ (whereas $e^K \to |\tau|^2\,e^K$) under $\tau \to -\frac{1}{\tau}$ while $(W \to W, K \to K)$ under $\tau \to \tau + 1$. These observations ensure that the subsequent $F$-term contributions to the scalar potential are invariant under $\tau \to -\frac{1}{\tau}$ as well as $\tau \to \tau +1$. From now onwards we will mainly focus on transformation $\tau \to -\frac{1}{\tau}$ which involves the strong/weak S-duality, however at some point in the concluding section we will comment on the effects of $\tau \to \tau +1$ as well.

\section{Invoking new generalized flux-orbits}
\label{sec_invokingOrbits}
We will now search for a new set of flux orbits which are generalized by the presence of odd-axions $B_2/C_2$, and the S-dual $P$-flux. This will generalize the results of \cite{Blumenhagen:2015kja,Gao:2015nra,Shukla:2015bca}. For that purpose, we divide our intuitive search into two parts; first we will invoke the new flux-orbits with odd-index $k\in h^{2,1}_-(X)$ and later with even-index $K\in h^{2,1}_+(X)$.  

\subsection{Flux orbits with odd-index $k\in h^{2,1}_-(X)$}
As flux-components with odd-index $k$ are involved in F-term contribution via the superpotential, let us rearrange the following three-form factor appearing in eqn. (\ref{eq:W2}),
\bea
& & \hskip-1.75cm \left({F} +\tau {H}\right) + \omega_a {G}^a + \left(\hat{Q}^{\alpha}+ \tau \hat{P}^{\alpha}\right){T}_\alpha -\hat{P}^{\alpha} \left(\frac{1}{2} \hat{\kappa}_{\alpha a b} G^a G^b\right)\\
& &  =\biggl[\left({\cal F}^k + s\, \hat{\cal P}^{\alpha\,k} \sigma_\alpha\right)  + i \, \left(s \, {\cal H}^k -\hat{{\cal Q}}^{\alpha\,k} \sigma_\alpha \right) \biggr]\, {\cal A}_k \, \nonumber\\
& & \quad \quad \quad + \biggl[\left({\cal F}_k + s\, \hat{\cal P}^{\alpha}{}_{k} \sigma_\alpha\right)  + i \, \left(s \, {\cal H}_k -\hat{{\cal Q}}^{\alpha}{}_{k} \sigma_\alpha \right) \biggr]\, {\cal B}^k \,, \nonumber
\eea
where $\sigma_\alpha = \frac{1}{2}\, {\kappa}_{\alpha \beta \gamma} t^\beta t^\gamma$ in Einstein-frame, along with the following new flux-combinations which generalize the Type IIB orientifold results of \cite{Gao:2015nra} with the inclusion of odd axions ($B_2/C_2$) and S-dual $P$-fluxes, 
\begin{subequations}
\bea
\label{eq:orbitsB1}
& \hskip-1.5cm {\cal H}_k =  {\bf h}_k~, \, \, \, \, \, &\hat{\cal Q}^{\alpha}{}_{k} ={ \bf{\hat{q}^{\alpha}{}_{k} }} + C_0 \, {\bf \hat{p}^{\alpha}{}_{k}} ~, \nonumber\\
& {\cal F}_k=   {\bf f}_k + C_0 \, ~{\bf h}_k~, \, \, \, \, \, &\hat{\cal P}^{\alpha}{}_{k} = {\bf \hat{p}^{\alpha}{}_{k}}~; \nonumber\\
& & \\
& \hskip-1.5cm {\cal H}^k =  {\bf h}^k~, \, \, \, \, \, &{\hat{\cal Q}}^{\alpha k} = {\bf \hat{q}^{\alpha k}} + C_0 \, {\bf \hat{p}^{\alpha k}} ~, \nonumber\\
& {\cal F}^k=   {\bf f}^k + C_0 \, ~{\bf h}^k~, \, \, \, \, \, &{\hat{\cal P}}^{\alpha k} = {\bf {\hat{p}}^{\alpha k}}~, \nonumber
\eea
where
\bea
\label{eq:orbitsB2}
& &  {\bf h}_k = H_k + (\omega_{ak} \, {b}^a) + \hat{Q}^\alpha{}_k \, \left(\frac{1}{2}\, \hat{\kappa}_{\alpha a b} b^a b^b\right)  + \hat{P}^\alpha{}_k \, \left({\rho}_\alpha\right) \nonumber\\
& &  {\bf f}_k = F_k + (\omega_{ak} \, {c}^a) - \hat{P}^\alpha{}_k \, \left(\frac{1}{2}\, \hat{\kappa}_{\alpha a b} c^a c^b\right) + \hat{Q}^\alpha{}_k \, \left({\rho}_\alpha + \hat{\kappa}_{\alpha a b} c^a b^b\right) \nonumber\\
&& \\
& &  {\bf h}^k = H^k + (\omega_{a}{}^{k} \, {b}^a) + \hat{Q}^{\alpha k} \, \left(\frac{1}{2}\, \hat{\kappa}_{\alpha a b} b^a b^b\right)  +\hat{P}^{\alpha k} \, \left({\rho}_\alpha \right)\nonumber\\
& &  {\bf f}^k = F^k + (\omega_a{}^k \, {c}^a) -  \hat{P}^{\alpha k} \, \left(\frac{1}{2}\, \hat{\kappa}_{\alpha a b} c^a c^b\right) + \hat{Q}^{\alpha k} \, \left({\rho}_\alpha + \hat{\kappa}_{\alpha a b} c^a b^b\right)\, \, \nonumber \\
& & {\bf \hat{q}^{\alpha}{}_{k}} = \hat{Q}^{\alpha}{}_{k} , \quad {\bf \hat{q}^{\alpha k}} = \hat{Q}^{\alpha k}, \nonumber\\
& & {\bf \hat{p}^{\alpha}{}_{k}} = \hat{P}^{\alpha}{}_{k}, \quad {\bf {\hat{p}}^{\alpha k}} = {\hat{P}}^{\alpha k} \, .\nonumber
\eea
\end{subequations}
{\it The flux components written in bold and small symbols are what we call new-generalized flux orbits as these follow the same strong/weak S-duality rules as to those of respective usual (non-)geometric fluxes along with the standard $H_3/F_3$ fluxes. Further, the flux components written in bold and capital symbols are the ones relevant for a `suitable' rearrangement of scalar potential as we will see in explicit examples later on.  We call that rearrangement `suitable' as the same might be useful for guessing the ten-dimensional origin of scalar potential via a dimensional oxidation process on the lines of \cite{Blumenhagen:2013hva,Gao:2015nra,Shukla:2015bca}.} 

To appreciate the structure of {\it new generlaized flux orbits} given in eqns. (\ref{eq:orbitsB1})-(\ref{eq:orbitsB2}), it is worth to point out the following observations,
\begin{itemize}
\item{In the absence of odd $(B_2/C_2)$-axions, our flux orbits reduce to the ones proposed in \cite{Gao:2015nra} while performing dimensional oxidation with the inclusion of P-flux.}
\item{Similar to type IIA orientifold case \cite{Blumenhagen:2013hva}, the $H_3$ flux receives corrections of type $(\omega \triangleleft B_2)$ and $Q\triangleright (B_2\wedge B_2)$. Moreover, the RR flux $F_3$ is also receiving corrections of type $(\omega \triangleleft C_2)$ and $P\triangleright (C_2 \wedge C_2)$ making the overall structure symmetric under strong/weak duality ! However, no correction of $R\bullet (B_2 \wedge B_2 \wedge B_2)$ or $R\bullet (C_2 \wedge C_2 \wedge C_2)$-type appears in the new ${\bf h}/{\bf f}$-flux orbits as $\hat{k}_{abc}=0$, by orientifold construction itself.}
\item{As expected, these new orbits generalize the results of \cite{Shukla:2015bca} obtained without including P-flux. Subsequently, we find that the pairs (${h}_k, {f}_k$) and (${h}^k, {f}^k$) are S-dual pairs, and the same is more obvious when one takes into account the fact that $\rho_\alpha \to \left({\rho}_\alpha + \hat{\kappa}_{\alpha a b} c^a b^b\right)$ under strong/weak S-duality. Also, if one wants to see a more symmetric version of eqn. (\ref{eq:orbitsB2}), one has to consider a S-duality invariant version of RR-four form, defined via $\tilde{\rho}_\alpha \equiv \rho_\alpha + \frac{1}{2}\hat{\kappa}_{\alpha a b} c^a b^b$. Subsequently, we have the following rearrangement of orbits which make the S-dual symmetry quite obvious,
\bea
\label{eq:orbitsB3}
& &  {\bf h}_k = H_k + (\omega_{ak} \, {b}^a) + \hat{Q}^\alpha{}_k \, \left(\frac{1}{2}\, \hat{\kappa}_{\alpha a b} b^a b^b\right)  + \hat{P}^\alpha{}_k \, \left(\tilde{\rho}_\alpha -\frac{1}{2} \hat{\kappa}_{\alpha a b} c^a b^b\right) \nonumber\\
& &  {\bf f}_k = F_k + (\omega_{ak} \, {c}^a) - \hat{P}^\alpha{}_k \, \left(\frac{1}{2}\, \hat{\kappa}_{\alpha a b} c^a c^b\right) + \hat{Q}^\alpha{}_k \, \left(\tilde{\rho}_\alpha +\frac{1}{2} \hat{\kappa}_{\alpha a b} c^a b^b\right)\, , 
\eea
and 
\bea
\label{eq:orbitsB4}
& &  {\bf h}^k = H^k + (\omega_{a}{}^{k} \, {b}^a) + \hat{Q}^{\alpha k} \, \left(\frac{1}{2}\, \hat{\kappa}_{\alpha a b} b^a b^b\right)  + \hat{P}^{\alpha k} \, \left(\tilde{\rho}_\alpha -\frac{1}{2} \hat{\kappa}_{\alpha a b} c^a b^b\right) \nonumber\\
& &  {\bf f}^k = F^k + (\omega_{a}{}^{k} \, {c}^a) - \hat{P}^{\alpha k} \, \left(\frac{1}{2}\, \hat{\kappa}_{\alpha a b} c^a c^b\right) + \hat{Q}^{\alpha k} \, \left(\tilde{\rho}_\alpha +\frac{1}{2} \hat{\kappa}_{\alpha a b} c^a b^b\right). 
\eea
}
\end{itemize}
Moreover, from type IIA-orientifold results of \cite{Blumenhagen:2013hva} and type IIB-orientifold results of \cite{Shukla:2015bca}, one observes that the geometric flux orbit can be corrected via ($Q\triangleright B_2$) and $R\bullet(B_2\wedge B_2$)-type of terms, and so we conjecture the following S-dual completion:
\bea
\label{eq:OddOrbitB5}
& & {\cal \mho}_{ak}\equiv {\bf \omega_{ak}} = \biggl[\omega_{ak} + \hat{Q}^\alpha{}_k \, \left(\hat{\kappa}_{\alpha a b}\, b^b\right) -\, \hat{P}^\alpha{}_k \, \left(\hat{\kappa}_{\alpha a b}\, c^b\right) \biggr]\\
& & {\cal \mho}_{a}{}^{k} \equiv {\bf \omega_{a}{}^{k}}= \biggl[\omega_{a}{}^{k} + \hat{Q}^{\alpha k} \, \left(\hat{\kappa}_{\alpha a b}\, b^b\right)-\hat{P}^{\alpha k} \, \left(\hat{\kappa}_{\alpha a b}\, c^b\right) \biggr]\nonumber
\eea
There are no $k$-index $R$-flux components allowed by the full orientifold action, and so we do not have any presence of $R$-flux in the flux orbits ${\cal \mho}_{ak}$ and ${\cal \mho}_{a}{}^{k}$. This orbit (\ref{eq:OddOrbitB5}), along with the previous S-dual pieces for NS-NS and RR flux orbits given in eqns. (\ref{eq:orbitsB1}-\ref{eq:orbitsB2}) and (\ref{eq:orbitsB3}-\ref{eq:orbitsB4}), completes our search for invoking the generalized components of various flux-orbits involving the odd-$H^{(2,1)}(X)$ cohomology indices, i.e. $k \in h^{2,1}_-(X)$.

\subsection{Flux orbits with even-index $K\in h^{2,1}_+(X)$}
Now we switch towards seeking for the strong/weak S-duality invariant version of the piece $V_D^{(1)}$. It has been observed in \cite{Shukla:2015bca} that looking at the rearrangement of such D-terms appearing in a setup with $h^{2,1}_+(X) \ne 0$ has helped in guessing the corrections to the geometric flux orbits with the inclusion of odd-axions very similar to the case of type IIA-orientifolds \cite{Blumenhagen:2013hva}. Let us recall that in the absence of P-fluxes, and using generic expressions for K\"ahler derivatives, the D-terms (\ref{eq:DtermOld}) can be rewritten as under \cite{Shukla:2015bca},
\bea
\label{eq:DtermOld1}
& & \hskip-1cm D_K  = \frac{f^{-1} R_K}{2\, s}\, - \frac{t^\alpha \,\hat{\mho}_{\alpha K}}{2 \, {\cal V}_E} \, , \, \, \, \,D^K  = -\frac{f^{-1} R^K}{2\, s}\, + \frac{t^\alpha \,\hat{\mho}_{\alpha}{}^{K}}{2 \, {\cal V}_E}.
\eea
where
\bea
\label{eq:mhoOld}
& & {\hat{\mho}_{\alpha K}} = \hat{\omega}_{\alpha K}\, - (d^{-1})_b{}^a  \, Q^{b}{}_{K} \, \, \left(\hat{k}_{\alpha a c} \, b^c \right) + f^{-1} \, \, R_K \, \left(\frac{1}{2}\hat{k}_{\alpha a b} \, b^a \,b^b\right) \\
& & {\hat{\mho}_{\alpha}{}^{K}} =\hat{\omega}_{\alpha}{}^{K}\, - (d^{-1})_b{}^a \, \, Q^{b K} \, \left(\hat{k}_{\alpha a c} \, b^c\right)+ f^{-1} \, \, R^K \, \left(\frac{1}{2}\hat{k}_{\alpha a b} \, b^a \,b^b\right) \nonumber
\eea
Now, we will provide a modular completion of all those geometric and non-geometric flux orbits counted via $h^{2,1}_+(X)$ indices, and with the inclusion of P-flux. 

Given that two-cycle volume moduli $t^\alpha$ as well as internal volume ${\cal V}_E$ are expressed in Einstein-frame and therefore are S-duality invariant, and so the second pieces in each of the $D$-terms given by eqn. (\ref{eq:DtermOld1}) along with (\ref{eq:mhoOld}) suggest the following modular completion of geometric flux orbits, 
\bea
& & {\bf{\hat{{\mathbb\mho}}_{\alpha K}}}\equiv {\bf \hat{\omega}_{\alpha K}} = \hat{\omega}_{\alpha K}\, - (d^{-1})_b{}^a  \, Q^{b}{}_{K} \, \, \left(\hat{k}_{\alpha a c} \, b^c \right) + f^{-1} \, \, R_K \, \left(\frac{1}{2}\hat{k}_{\alpha a b} \, b^a \,b^b\right)\nonumber\\
& & \hskip2.4cm + (d^{-1})_b{}^a  \, P^{b}{}_{K} \, \, \left(\hat{k}_{\alpha a c} \, c^c \right) + f^{-1} \, \, R_K \, \left(\frac{1}{2}\hat{k}_{\alpha a b} \, c^a \,c^b\right)\\
& & {\bf \hat{{\mathbb\mho}}_{\alpha}{}^{K}} \equiv {\bf \hat{\omega}_{\alpha}{}^{K}} =\hat{\omega}_{\alpha}{}^{K}\, - (d^{-1})_b{}^a \, \, Q^{b K} \, \left(\hat{k}_{\alpha a c} \, b^c\right)+ f^{-1} \, \, R^K \, \left(\frac{1}{2}\hat{k}_{\alpha a b} \, b^a \,b^b\right)\nonumber\\
& & \hskip2.4cm + (d^{-1})_b{}^a \, \, P^{b K} \, \left(\hat{k}_{\alpha a c} \, c^c\right)+ f^{-1} \, \, R^K \, \left(\frac{1}{2}\hat{k}_{\alpha a b} \, c^a \,c^b\right)\nonumber
\eea
while the modular completion of first piece with $R$-flux in eqn. (\ref{eq:DtermOld1}) is tricky. One observes that $R$-flux piece qualitatively appears as $|R|^2/s^2$ in respective D-term scalar potential. Now given that $R$-flux is S-duality invariant, in order to have an overall modular invariance, one needs to add a piece of kind $(|\tau|^4 \times |R|^2)/s^2$. This dilaton dependence of kind $s^{-2}$ in the prefactor of (Einstein frame\footnote{Note that, all the quadratic flux pieces involving NS-NS fluxes $H, \omega, Q$ and $R$ appear with the same `explicit' dilaton dependence $e^{-2 \phi}$ \cite{Blumenhagen:2013hva}, however when changed into Einstein frame, the overall effective dilaton factors change due to their appearances inside metric- and inverse-metric components involved in the contraction within a particular term \cite{Gao:2015nra,Shukla:2015bca}.}) $R^2$-term is unlike the other flux-squared pieces for which it is always of the type $s^a$ with $a \in \{-1, 0 ,1\}$ only. More specifically, for ${\cal F}{\cal F}, {\cal F}{\cal Q},{\cal Q}{\cal Q}$ and ${R}{\cal \mho}$ pieces, we have $a = -1$ while for ${\cal H}{\cal H}, {\cal H}{\cal P}$ and ${\cal P}{\cal P}$ pieces, we have $a = 1$. All other flux-squared terms (e.g. ${\cal H}{\cal Q}, {\cal F}{\cal P}, {\mho}{\mho}, {\cal P}{\cal Q}$-types) do not have dilaton dependence in Einstein frame. Therefore, {\it one possibility} for the invariance of D-term scalar potential under strong/weak S-duality can be anticipated as under,
\bea
\label{eq:newD-term}
& & \hskip-1cm {\bf D_K}  = \frac{f^{-1} {\bf R}_K}{2}\, - \frac{t^\alpha \,{\bf \hat{\mho}_{\alpha K}}}{2 \, {\cal V}_E} \, , \, \, \, \,{\bf D^K}  = -\frac{f^{-1} {\bf R}^K}{2}\, + \frac{t^\alpha \,{\bf \hat{\mho}_{\alpha}{}^{K}}}{2 \, {\cal V}_E},
\eea
where 
\bea
{\bf R}_K \equiv {\bf r_K}= \frac{(1+|\tau|^2)}{s}\, R_K\, , \quad \quad \quad {\bf R}^K \equiv {\bf r^K}= \frac{(1+|\tau|^2)}{s}\, R^K.
\eea
Recalling type IIA orientifold results of \cite{Blumenhagen:2013hva} and type IIB orientifolds results of \cite{Shukla:2015bca}, we have seen that the geometric flux orbits are corrected via ($Q\triangleright B_2$) and $R\bullet(B_2\wedge B_2$)-type of terms. Moreover, corrections of types $(R\bullet B_2)$ are indeed expected in the even-index $Q$-flux components though the same was not possible for odd indexed ones because of orientifold projection. Therefore, one expects that Q-flux and P-flux orbits are generalized in the following manner to respect the strong/weak S-duality symmetry,
\bea
\label{eq:EvenOrbitA}
& & {\bf q}^{a{}K} ={Q}^{a{}K} + f^{-1} \,\, d_b{}^a  \, (R^K \bullet \, b^b)\,, \quad \quad {\bf q}^{a}{}_{K} = {Q}^{a}{}_{K} + f^{-1} \,\,d_b{}^a \, (R_K \bullet b^b) \\
& & {\bf p}^{a{}K} ={P}^{a{}K} - f^{-1} \,\, d_b{}^a  \, (R^K \bullet \, c^b)\,, \quad \quad {\bf p}^{a}{}_{K} = {P}^{a}{}_{K} - f^{-1} \,\,d_b{}^a \, (R_K \bullet c^b) \nonumber
\eea
Therefore, similar to eqn. (\ref{eq:orbitsB1}), the structure to appear in the rearrangement of the potential can be anticipated as,
\bea
\label{eq:orbitsB7}
& & {\cal Q}^{a{}K}  = {\bf q}^{a{}K}  + C_0 \, {\bf p}^{a{}K} ~, \, \quad {\cal Q}^{a}{}_{K}  = {\bf q}^{a}{}_{K} + \, C_0 {\bf p}^{a}{}_{K}~, \\
& & {\cal P}^{a{}K}  = {\bf p}^{a{}K} ~, \, \quad \hskip2cm  {\cal P}^{a}{}_{K}  = {\bf p}^{a}{}_{K} \, .\nonumber
\eea
Before we come to explicit examples, let us emphasize that our new generalized flux orbits respect the same strong/weak-duality as to those of the older generalized fluxes in the following manner,
\bea
& & {\bf h_k} \to  {\bf f_k} , \quad  {\bf h^k} \to  {\bf f^k}, \quad \quad \quad  \quad \quad \quad \quad {\bf f_k} \to - {\bf h_k} , \quad  {\bf f^k} \to - {\bf h^k} \nonumber\\
& &  {\bf q^a_k} \to - {\bf p^a_k} , \quad  {\bf q^{ak}} \to - {\bf p^{ak}}, \quad \quad \quad \quad {\bf p^a_k} \to  {\bf q^a_k} , \quad  {\bf p^{ak}} \to  {\bf q^{ak}}\\
& &  {\bf\hat{q}^\alpha_K} \to - {\bf\hat{p}^\alpha_K} , \quad  {\bf\hat{q}^{\alpha K}} \to - {\bf\hat{p}^{\alpha K}}, \quad \quad \quad  {\bf\hat{p}^\alpha_K} \to  {\bf\hat{q}^\alpha_K} , \quad  {\bf\hat{p}^{\alpha K}} \to  {\bf\hat{q}^{\alpha K}} \nonumber
\eea
while all the components of {\it new} geometric flux, namely (${\bf {\omega}_{ak}}, {\bf {\omega}_{a}{}^{k}}, {\bf{\hat{{\omega}}_{\alpha K}}}, {\bf \hat{{\omega}}_{\alpha}{}^{K}}$) as well as the {\it new} non-geometric flux ${\bf r}_K, {\bf r}^K$ are internally self-dual similar to the usual geometric and non-geometric fluxes.

We will use these pieces of information about the {\it new generalized flux orbits} for collecting a compact rearrangement of the total $F$-term contributions. {\it Further, it is manifested that the F-term coming from the superpotential (\ref{eq:W2}) is S-duality invariant by construction itself, which we also verify in our explicit computations of two concrete models}.

\section{Scalar potential rearrangements in two concrete examples}
\label{sec_rearrangement}
To illustrate the utility of our new flux-orbits, in this section, we consider two concrete type IIB orientifold examples with frozen complex structure moduli. Explicit construction of toroidal setups with frozen complex structure moduli can be found in \cite{Lust:2006zh}. Our current aim is to rewrite the four dimensional scalar potential by using our new generalized flux orbits.

\subsection{Model A}
In this example, we consider the compactifying CY to have $h^{2,1}(X) = 1_+ +0_-$, and $h^{1,1}(X) = 1_+ +1_-$. The same implies that this setup has one complex K\"ahler modulus ($T$), one odd axion ($G$) and no complex structure moduli. Therefore, the bosonic part of resulting 4D effective theory can be described by axion-dilaton $\tau$, one $T$ and one $G$ modulus.  %For constructions with no-complex structure moduli, see \cite{}. 
Now the K\"ahler potential can be given as,
\bea
& &\hskip-0.05cm  K = - \ln\left(-i(\tau-\ov\tau)\right) -3\ln\biggl[i\left(T-\ov T -\frac{\hat{\kappa} (G-\ov G)^2}{2(\tau-\ov \tau)}\,\right)\biggr], \nonumber
\eea
where appropriate normalization $\left(i\int_{X}\Omega_3\wedge{\bar\Omega_3}\right) = 1$ has been made via considering $\Omega_3 = \frac{1}{\sqrt2}({\cal A}_0 + i\, {\cal B}^0 )$. Further, we have the only non-zero flux components with indices $k=0$ and $K=1$ which are given as,
\bea
& & \hskip-0.5cm {k}: \quad F_0, F^0, \, H_0, H^0,\,  {\omega}{}^0, {\omega}_{0}, \, \hat{Q}^{{}0} , \, \hat{Q}^{}{}_{0} , \, \hat{P}^{{}0} , \, \hat{P}^{}{}_{0} \nonumber\\
& & \hskip-0.5cm {K}: \quad \hat{\omega}{}^1, \hat{\omega}_{1}, \, {Q}^{{}1}, \, {Q}^{}{}_{1}, \, {P}^{{}1}, \, {P}^{}{}_{1}, \, R_1, R^1\, .
\eea
Here indices $\alpha$ and $a$ counting $T_\alpha$ and $G^a$ moduli are dropped. The simplified form of superpotential becomes,
\bea
& & \hskip-1.5cm W = \frac{1}{\sqrt2} \biggl[\left(F_0 +\tau H_0 +{\omega}_{0} G + (\hat{Q}^{}{}_{0} +\tau \hat{P}^{}{}_{0} ) T -\hat{P}^{}{}_{0} \frac{\hat{\kappa} \, \, G^2}{2} \right) \\
& & \quad - i \, \, \left(F^0 +\tau H^0 +{\omega}^{0} G + (\hat{Q}^{}{}^{0} +\tau \hat{P}^{}{}^{0} ) T -\hat{P}^{}{}^{0} \frac{\hat{\kappa} \, \, G^2}{2} \right) \biggr] . ~\nonumber
\eea 
The new flux orbit are given as,
\bea
\label{eq:orbitsB1Ex1}
&& {\cal H}_0 =  {\bf h}_0~, \, \, \, \, \, \hat{\cal Q}_{0} ={ \bf{\hat{q}_{0} }} + C_0 \, {\bf \hat{p}_{0}} ~, \quad \quad  {\cal F}_0=   {\bf f}_0 + C_0 \, ~{\bf h}_0~, \, \, \, \, \, \hat{\cal P}_{0} = {\bf \hat{p}_{0}}, \nonumber
%& & \\
%&& {\cal H}^0 =  {\bf h}^0~, \, \, \, \, \, {\hat{\cal Q}}^{0} = {\bf \hat{q}^{0}} + C_0 \, {\bf \hat{p}^{ 0}} ~, \nonumber\\
%&& {\cal F}^0=   {\bf f}^0 + C_0 \, ~{\bf h}^0~, \, \, \, \, \, {\hat{\cal P}}^{0} = {\bf {\hat{p}}^{ 0}}~, \nonumber
\eea
where
\bea
\label{eq:orbitsB2Ex1}
& & \hskip-0.5cm {\bf h}_0 = H_0 + \omega_{0} \, {b} + \hat{Q}_0 \, \left(\frac{1}{2}\, \hat{\kappa} \, b^2\right)  +  \hat{P}{}_0 \,{\rho}, \quad \quad {\cal \mho}_{0} = \omega_{0} + \hat{Q}_0 \, \hat{\kappa}\, b -\, \hat{P}_0 \, \hat{\kappa}\, c\, ,  \\
& & \hskip-0.5cm  {\bf f}_0 = F_0 + \omega_{0} \, {c}  - \hat{P}_0 \, \left(\frac{1}{2}\, \hat{\kappa} \, c^2 \right) + \hat{Q}_0 \,\left({\rho} + \hat{\kappa}\, b \, c \right), \quad {\bf \hat{q}_{0}} = \hat{Q}_{0} , \quad  {\bf \hat{p}_{0}} = \hat{P}_{0}. \nonumber
\eea
Similarly all the other components with upper index being `$k=0$' can also be written. The resulting F-term scalar potential has 192 terms in total, and using our new flux-orbits, on the lines of \cite{Blumenhagen:2013hva,Gao:2015nra,Shukla:2015bca} we can have the following `suitable' rearrangement,
\bea
\label{eq:Voddflux}
& & \hskip-0.8cm  V_{{\cal F}{\cal F}} =  \frac{{\cal F}_0^2+{({\cal F}^0)}^2 }{4 \, s\,{\cal V}_E^2}, \, V_{{\cal H}{\cal H}} =  \frac{s\, {\cal H}_0^2 +s\, {({\cal H}^0)}^2 }{4 \, {\cal V}_E^2}, \, V_{\hat{\cal Q}\hat{\cal Q}} = -5\, \frac{\hat{\cal Q}_0^2+{(\hat{\cal Q}^0)}^2}{3 \, s\, {\cal V}_E^{2/3}}\\
& & \hskip-0.8cm V_{{\cal \mho}{\cal \mho}} = -\, \frac{{\cal \mho}_0^2+{({\cal \mho}^0)}^2}{3 \,\hat{\kappa} \, \, {\cal V}_E^{4/3}} \, , \,V_{{\cal H}\hat{\cal Q}} = 3\, \frac{{\cal H}_0 \hat{\cal Q}_0+\hat{\cal Q}^0 {\cal H}^0 }{2 \, {\cal V}_E^{4/3}} , \, V_{{\cal F}\hat{\cal P}} = - 3\, \frac{{\cal F}_0 \hat{\cal P}_0+\hat{\cal P}^0 {\cal F}^0 }{2 \, {\cal V}_E^{4/3}}, \nonumber\\
& & \hskip-0.8cm  V_{\hat{\cal P}\hat{\cal P}} = - 5\, \frac{s\, \hat{\cal P}_0^2 +s\, {(\hat{\cal P}^0)}^2}{3 \, {\cal V}_E^{2/3}} \, , \, \, V_{\hat{\cal P}\hat{\cal Q}} =  11\frac{\hat{\cal P}_0 \hat{\cal Q}^0-\hat{\cal Q}_0 \hat{\cal P}^0 }{3 \, {\cal V}_E^{2/3}},\nonumber\\
& & \hskip-0.8cm V_{{\cal F}{\cal H}} =  \frac{{\cal H}_0 {\cal F}^0-{\cal F}_0 {\cal H}^0 }{2 \, {\cal V}_E^2}, \, \, V_{{\cal F}\hat{\cal Q}} =  \frac{{\cal F}_0 \hat{\cal Q}^0-\hat{\cal Q}_0 {\cal F}^0 }{2 \,s\, {\cal V}_E^{4/3}} , \, V_{{\cal H}\hat{\cal P}} =  \frac{s\,{\cal H}_0 \hat{\cal P}^0-s\,\hat{\cal P}_0 {\cal H}^0 }{2 \, {\cal V}_E^{4/3}}. \nonumber
\eea
The counting of terms in these pieces goes as: $\#(V_{{\cal H}{\cal H}}) = 20,  \#(V_{{\cal F}{\cal F}}) = 88, \#(V_{{\cal F}{\cal H}}) = 30, \#(V_{{\cal F}{\cal Q}}) = 22, \#(V_{{\cal H}{\cal P}}) = 6, \#(V_{{\cal P}{\cal Q}}) = 2, \#(V_{{\cal Q}{\cal Q}}) = 6, \#(V_{{\cal P}{\cal P}}) = 2$ while $\#(V_{{\cal \mho}{\cal \mho}}+V_{{\cal H}{\cal Q}}+V_{{\cal F}{\cal P}}) = 16$, and everything sums to a total of 192 terms to have a perfect match with F-term contribution. Note that, similar to the case of type IIB superstring theory compactified on a ${\mathbb T}^6/{\mathbb Z}_4$-orientifold \cite{Shukla:2015bca}, in order to see the piecewise counting of terms, one has to consider the combination $(V_{{\cal \mho}{\cal \mho}}+V_{{\cal H}{\cal Q}}+V_{{\cal F}{\cal P}})$ which involves internal cancellations. By saying this we mean that though individual pieces among $V_{{\cal \mho}{\cal \mho}}, V_{{\cal H}{\cal Q}}$ and $V_{{\cal F}{\cal P}}$ have some terms which are not present in $F$-term contributions, however once we consider the sum $V_{{\cal \mho}{\cal \mho}}+V_{{\cal H}{\cal Q}}+V_{{\cal F}{\cal P}}$, all terms are indeed part of $F$-term scalar potential.

In addition, as seen from the structure of D-terms (\ref{eq:newD-term}), the contributions to 4D effective potential are of ($R^2, \hat{\mho} R, \hat{\mho}^2$)-types very analogous to type IIA case \cite{Blumenhagen:2013hva}. 
\bea
\label{eq:D-termExampleA}
& & \hskip-1.5cm V_{{\cal R}{\cal R}} = \frac{{\cal R}_1^2+({\cal R}^1)^2}{4 \,f^2} \equiv \#12, \, V_{{\hat{\cal \mho}}{\hat{\cal \mho}}} =  \frac{\hat{{\cal \mho}}_1^2+{({\hat{\cal \mho}^1})}^2 }{4\, \,{\cal V}_E^{4/3}}\, \equiv \#30, \nonumber\\
& & \hskip-0.04cm V_{{\cal R}{\hat{\cal \mho}}}=-\, \frac{{\cal R}_1 {\hat{\cal \mho}}_1+ {\cal R}^1 \, {\hat{\cal \mho}}^1}{2 \,f\, {\cal V}_E^{2/3}} \equiv \#30\, , % \, \, \, {\cal R}_1 = r_1 \,, \, \, {\cal R}^1 = r^1. 
\eea
where
\bea
& & {\bf{\hat{{\mathbb\mho}}_{1}}}\equiv {\bf \hat{\omega}_{1}} = \hat{\omega}_{1}\, -  Q_{1} \, \, \left(\hat{\kappa} \, b \right) + R_1 \, \left(\frac{1}{2}\hat{\kappa} \, b^2\right) + \, P_{1} \, \, \left(\hat{\kappa} \, c \right) + \, R_1 \, \left(\frac{1}{2}\hat{\kappa} \, c^2\right), \nonumber\\
& & {\cal R}_1 \equiv {\bf r_1}= \frac{(1+|\tau|^2)}{s}\, R_1\, 
\eea
and similarly the flux components with upper index can be written. It is quite interesting to see the $R$-flux squared term appearing via D-term contributions in a similar fashion. Recall that the same has been not allowed via F-term because orientifold projection does not allow R-flux to appear in the superpotential. Thus, these rearrangement of F/D-terms using our new flux-orbits may suggest some important features for understanding 10D non-geometric S-duality invariant action.

Let us make the following two points about the use of this `suitable' rearrangement of the scalar potential,
\begin{itemize}
\item{The flux-squared pieces with both indices of same-types lower/upper as seen in eqns. (\ref{eq:Voddflux}) and (\ref{eq:D-termExampleA}), e.g.  (${\cal X}_0  {\cal Y}_0 + {\cal X}^0  {\cal Y}^0$) can be understood to be originated from $({\cal X} \wedge \ast {\cal Y})$-type in the ten dimensional perspective.}
\item{The scalar potential pieces with one index up and other down, e.g. $({\cal X}_0  {\cal Y}^0-{\cal X}_0  {\cal Y}^0)$ can appear via $({\cal X} \wedge {\cal Y})$-type terms in ten-dimensional version. }
\end{itemize}
So, up to fixing the coefficients correctly, the form of ten-dimensional pieces which could give rise to the total scalar potential after dimensional reduction can be intuitively guessed \footnote{The structure of each of the flux-squared pieces, except the ones involving $P$-flux, in eqns. (\ref{eq:Voddflux}) and (\ref{eq:D-termExampleA}) of scalar potential rearrangement could be later derived in a symplectic formulation of the scalar potential  \cite{Shukla:2015hpa} using the same generalized flux orbits proposed in this article.}.

\subsection{Model B}
In this case, we consider a type IIB compactification setup on the orientifold of ${\mathbb T}^6/{\mathbb Z}_4$ orbifold and analyze the scalar potential for the untwisted sector moduli/axions. This setup has $h^{2,1}(X) = 1_+ +0_-$, and $h^{1,1}(X) = 3_+ +2_-$, i.e. there are three complexified K\"ahler moduli ($T_\alpha$), two complexified odd axions ($G^a$) and no complex structure moduli. While we leave all the orientifold construction related details to be directly referred from \cite{Shukla:2015bca,Robbins:2007yv}, here we will simply provide the explicit expressions of K\"ahler- and super-potentials for analyzing F-term scalar potential. The K\"ahler potential is given as under,
\bea
\label{typeIIBK}
K &= -\ln\left(-i(\tau-\ov \tau)\right) - 2 \, \ln{\cal V}_E(T_\alpha, \tau, G^a; \ov T_\alpha, \ov \tau, \ov G^a)
\eea
where the Einstein frame volume is given as,
\bea
\label{eq:Volume}
& & {\cal V}_E\equiv {\cal V}_E(T_\alpha, S, G^a) = \frac{1}{4}\, \, \left(\frac{i(T_3 -\ov T_3)}{2} - \frac{i}{4 (\tau -\ov \tau)} \, \hat{\kappa}_{3 a b} \, (G^a -\ov G^a) (G^b -\ov G^b)\right)^{1/2} \nonumber\\
& & \hskip2cm \times  \biggl[\biggl(\frac{i(T_1 -\ov T_1)}{2}\biggr)^2 -2\, \biggl(\frac{i(T_2 -\ov T_2)}{2}\biggr)^2\biggr]^{1/2} \, \, 
\eea
%As it is reflected in the volume form, only $\hat\kappa_{3ab}$ intersection numbers are non-zero, and moreover $\hat\kappa_{3ab} = -4, \hat\kappa_{3ab} = -2$. 
Also, similar to the previous example, given that $h^{2,1}_-(X) =0$, complex structure moduli dependent piece of the K\"ahler potential is just a constant piece which can be nullified via choosing an appropriate normalization $\left(i\int_{X}\Omega_3\wedge{\bar\Omega_3}\right)=1$. Further, the generic form of the tree level flux-superpotential with all allowed fluxes being included is given as,
\bea
\label{eq:Wsimp1}
& & \hskip-1.5cm W = \frac{1}{\sqrt2}\biggl[\left(f_0 + \tau\, h_0 + \omega_{a0} \, G^a + \hat{Q}^{\alpha}{}_{0} \, T_\alpha + \tau\, \hat{P}^{\alpha}{}_{0} \, T_\alpha -\,\hat{P}^{\alpha}{}_{0} \left(\frac{1}{2} \hat{\kappa}_{\alpha a b} G^a G^b\right)\right) \nonumber\\
& & -i\,  \left(f^0 + \tau\, h^0 + \omega_a{}^0 \, G^a + \hat{Q}^{\alpha 0} \, T_\alpha +\tau\, \hat{P}^{\alpha 0} \, T_\alpha  -\,\hat{P}^{\alpha 0} \left(\frac{1}{2} \hat{\kappa}_{\alpha a b} G^a G^b\right)\right)\biggr]\,,
\eea
where $a = \{1,2\}$ and $\alpha=\{1,2,3\}$. Now, one can compute the full F-term scalar potential from these explicit expressions of $K$ and $W$. Let us mention that the explicit expressions for the new flux orbits are given as,
\begin{subequations}
\bea
\label{eq:orbitsB1Ex2}
& \hskip-1.4cm {\cal H}_0 =  {\bf h}_0~, \, \, \, \, \, &\hat{\cal Q}^{\alpha}{}_{0} ={ \bf{\hat{q}^{\alpha}{}_{0} }} + C_0 \, {\bf \hat{p}^{\alpha}{}_{0}} ~, \\
& {\cal F}_0=   {\bf f}_0 + C_0 \, ~{\bf h}_0~, \, \, \, \, \, &\hat{\cal P}^{\alpha}{}_{0} = {\bf \hat{p}^{\alpha}{}_{0}}~; \nonumber
%& & \\
%&& {\cal H}^0 =  {\bf h}^0~, \, \, \, \, \, {\hat{\cal Q}}^{\alpha 0} = {\bf \hat{q}^{\alpha 0}} + C_0 \, {\bf \hat{p}^{\alpha 0}} ~, \nonumber\\
%&& {\cal F}^0=   {\bf f}^0 + C_0 \, ~{\bf h}^0~, \, \, \, \, \, {\hat{\cal P}}^{\alpha 0} = {\bf {\hat{p}}^{\alpha 0}}~, \nonumber
\eea
where
\bea
\label{eq:orbitsB2Ex2}
& & \hskip-0.5cm {\bf h}_0 = H_0 + (\omega_{01} \, {b}^1+\omega_{02} \, {b}^2) + \hat{Q}^3{}_0 \, \left(\frac{1}{2}\, \hat{\kappa}_{3 11} (b^1)^2 + \frac{1}{2}\, \hat{\kappa}_{3 22} (b^2)^2\right)  +  \left(\hat{P}^1{}_0 \,{\rho}_1 + \hat{P}^2{}_0 \,{\rho}_2 + \hat{P}^3{}_0 \,{\rho}_3\right) \nonumber\\
& & \hskip-0.5cm  {\bf f}_0 = F_0 + (\omega_{01} \, {c}^1 + \omega_{02} \, {c}^2) - \hat{P}^3{}_0 \, \left(\frac{1}{2}\, \hat{\kappa}_{3 11} (c^1)^2+\frac{1}{2}\, \hat{\kappa}_{3 22} (c^2)^2 \right) \nonumber\\
& &  \hskip2cm + \hat{Q}^1{}_0 \,{\rho}_1 + \hat{Q}^2{}_0 \,{\rho}_2+ \hat{Q}^3{}_0\, \left({\rho}_3 + \hat{\kappa}_{3 11} c^1 b^1 + \hat{\kappa}_{3 22} c^2 b^2\right) \nonumber\\
%&& \\
& & \hskip-0.5cm {\cal \mho}_{01}\equiv {\bf \omega_{01}} = \biggl[\omega_{01} + \hat{Q}^3{}_0 \, \left(\hat{\kappa}_{311}\, b^1\right) -\, \hat{P}^3{}_0 \, \left(\hat{\kappa}_{311}\, c^1\right) \biggr] \\
& &  \hskip-0.5cm {\cal \mho}_{02}\equiv {\bf \omega_{02}} = \biggl[\omega_{02} + \hat{Q}^3{}_0 \, \left(\hat{\kappa}_{322}\, b^2\right) -\, \hat{P}^3{}_0 \, \left(\hat{\kappa}_{322}\, c^2\right) \biggr]\nonumber\\
%& &  {\bf h}^0 = H^0 + (\omega_{a}{}^{0} \, {b}^a) + \hat{Q}^{\alpha 0} \, \left(\frac{1}{2}\, \hat{\kappa}_{\alpha a b} b^a b^b\right)  +\hat{P}^{\alpha 0} \, \left({\rho}_\alpha \right)\nonumber\\
%& &  {\bf f}^0 = F^0 + (\omega_a{}^0 \, {c}^a) -  \hat{P}^{\alpha 0} \, \left(\frac{1}{2}\, \hat{\kappa}_{\alpha a b} c^a c^b\right) + \hat{Q}^{\alpha 0} \, \left({\rho}_\alpha + \hat{\kappa}_{\alpha a b} c^a b^b\right)\, \, \nonumber \\
& &  \hskip-0.5cm {\bf \hat{q}^{1}{}_{0}} = \hat{Q}^{1}{}_{0} , \quad {\bf \hat{q}^{2}{}_{0}} = \hat{Q}^{2}{}_{0}, \quad {\bf \hat{q}^{3}{}_{0}} = \hat{Q}^{3}{}_{0}, \quad \quad  {\bf \hat{p}^{1}{}_{0}} = \hat{P}^{1}{}_{0}, \quad {\bf \hat{p}^{2}{}_{0}} = \hat{P}^{2}{}_{0}, \quad {\bf \hat{p}^{3}{}_{0}} = \hat{P}^{3}{}_{0}% \quad {\bf {\hat{p}}^{\alpha 0}} = {\hat{P}}^{\alpha 0} 
\nonumber
\eea
\end{subequations}
and similarly all the other components with upper index being `$k=0$' can be written. Now, our aim is to show that using these new flux orbits, the huge $F$-term scalar potential, which results in a total of 960 terms, can be very compactly rearranged. To appreciate the structure as well as clarifying the step-by-step strategy to be followed, let us elaborate on the following three steps:
\subsubsection*{Step 1} %The following pieces are easiest to capture in the total $F$-term contributions. 
One can easily find the following explicit terms sitting within  the F-term pieces, and the same can also be cross-checked via a simple counting of terms mentioned with each piece as under,
\bea
\label{eq:5termsEx2}
& & \hskip-1cm V_{{\cal H}{\cal H}} =  \frac{s }{4 \, {\cal V}_E^2}\, \biggl[{\cal H}_0^2 +\, {({\cal H}^0)}^2 \biggr] \hskip4.9cm \#(V_{{\cal H}{\cal H}})=72 \, \nonumber\\
& & \hskip-1cm V_{{\cal F}{\cal F}} =  \frac{1}{4 \, s\,{\cal V}_E^2}\, \biggl[{\cal F}_0^2+{({\cal F}^0)}^2\biggr] \hskip4.9cm \#(V_{{\cal F}{\cal F}})=338\nonumber\\
& & \hskip-1cm V_{{\cal F}{\cal H}} =  \frac{1}{4 \, {\cal V}_E^2} \times 2 \biggl[{\cal H}_0 {\cal F}^0-{\cal F}_0 {\cal H}^0\biggr] \hskip4.0cm \#(V_{{\cal F}{\cal H}})=136 \nonumber\\
& & \hskip-1cm V_{{\cal F}\hat{\cal Q}} =  \frac{1}{4 \, s \, {\cal V}_E^2} \times 2\biggl[{\cal F}_0 \left(\hat{\cal Q}^{0\alpha}\sigma_\alpha \right) - \left(\hat{\cal Q}_0{}^\alpha \, \sigma_\alpha \right) {\cal F}^0 \biggr] \hskip1.3cm \#(V_{{\cal F}\hat{\cal Q}})= 180 \\
& & \hskip-1cm V_{{\cal H}\hat{\cal P}} =  \frac{s}{4 \, {\cal V}_E^2} \times 2 \biggl[{\cal H}_0 \left(\hat{\cal P}^{0\alpha}\sigma_\alpha \right) - \left(\hat{\cal P}_0{}^\alpha \, \sigma_\alpha \right) {\cal H}^0 \biggr] \hskip1.6cm \#(V_{{\cal H}\hat{\cal P}})=42 \nonumber
\eea
Thus, above five pieces capture 768 terms out of a total of 960 terms of F-term scalar potential, and subsequently only 192 terms are left. To appreciate the collections written out using our new generalized flux orbits, we have expressed the details of one simple piece ($V_{{\cal H}{\cal H}}$) in terms of original generalized fluxes in eqn. (\ref{eq:append1}) of the appendix A. %Here it is worth to recall again that similar to model A, last three pieces of eqn. (\ref{eq:5termsEx2}) correspond to 3/5/7-brane tadpoles subject to satisfying a set of Bianchi identities (\ref{eq:BIs2}). The same confirms the utility of our new generalized flux orbits.
\subsubsection*{Step 2} Unlike the case of ${\mathbb T}^6/({\mathbb Z}_2 \times {\mathbb Z}_2)$-orientifold \cite{Blumenhagen:2013hva}, due to a non-trivial mixing in our Model B with ${\mathbb T}^6/{\mathbb Z}_4$ setup, rest of the terms are not as nicely separated as those in step 1. Nevertheless one finds that a sum of the following three terms are precisely captured as 100 additional terms of $F$-term scalar potential.
\bea
\label{eq:5termsEx21}
& & \hskip-1cm V_{{\cal \mho}{\cal \mho}} = \frac{1}{4 \, {\cal V}_E^2} \biggl[\, \sigma_3\, \left(\mho_{01} \mho_{01} + \mho_{1}{}^{0} \mho_{1}{}^{0}\right) + 2\, \sigma_3 \left(\mho_{02} \mho_{02} + \mho_{2}{}^{0} \mho_{2}{}^{0}\right)\biggr]\nonumber\\
& & \hskip-1cm V_{{\cal H}\hat{\cal Q}} =  \frac{1}{4  \, {\cal V}_E^2} \times (+2)\biggl[3 \,{\cal H}^0 \left(\hat{\cal Q}^{0\alpha}\sigma_\alpha \right) + 3\, \left(\hat{\cal Q}_0{}^\alpha \, \sigma_\alpha \right) {\cal H}_0 \biggr]  \nonumber\\
& & \hskip-1cm V_{{\cal F}\hat{\cal P}} =  \frac{1}{4 \, {\cal V}_E^2} \times (-2) \biggl[3\, {\cal F}^0 \left(\hat{\cal P}^{0\alpha}\sigma_\alpha \right) + 3\, \left(\hat{\cal P}_0{}^\alpha \, \sigma_\alpha \right) {\cal F}^0 \biggr] \\
& & \hskip3cm \#(V_{{\mho}{\mho}} + V_{{\cal H}\hat{\cal Q}} + V_{{\cal F}\hat{\cal P}})=100 \nonumber
\eea
Note that, similar to Model A, the three pieces $V_{{\mho}{\mho}}, V_{{\cal H}\hat{\cal Q}}$ and $V_{{\cal F}\hat{\cal P}}$ individually have terms which are not present in F-term scalar potential, however after summing them up, some internal cancellation occurs to result in a total of 100 terms which are indeed among the additional ones in 192 terms. 
\subsubsection*{Step 3} After step 2, now one is left with rewriting just 92 terms of F-term scalar potential involving Q/P-fluxes, and the same can be done as under,
\bea
\label{eq:5termsEx22}
& & \hskip-1.0cm V_{\hat{\cal Q}\hat{\cal Q}} =\frac{1}{4 \,s\, {\cal V}_E^2} \biggl[\left(4 \sigma_2^2 - \sigma_1^2\right) \hat{\cal Q}_0{}^1 \hat{\cal Q}_0{}^1 + \left(\sigma_1^2 - \sigma_2^2\right) \hat{\cal Q}_0{}^2 \hat{\cal Q}_0{}^2  + \sigma_3^2 \hat{\cal Q}_0{}^3 \hat{\cal Q}_0{}^3 \nonumber\\
& & + 2 \, \sigma_1 \sigma_2 \hat{\cal Q}_0{}^1 \hat{\cal Q}_0{}^2 - 6 \, \sigma_2 \sigma_3 \hat{\cal Q}_0{}^2 \hat{\cal Q}_0{}^3- 6 \, \sigma_1 \sigma_3 \hat{\cal Q}_0{}^1 \hat{\cal Q}_0{}^3 \biggr] \hskip3.0cm \#(V_{\hat{\cal Q}\hat{\cal Q}})=54 \nonumber\\
& & \hskip-0.0cm + \frac{1}{4 \,s\, {\cal V}_E^2} \biggl[\left(4 \sigma_2^2 - \sigma_1^2\right) \hat{\cal Q}^{01} \hat{\cal Q}^{01} + \left(\sigma_1^2 - \sigma_2^2\right) \hat{\cal Q}^{02} \hat{\cal Q}^{02}  + \sigma_3^2 \hat{\cal Q}^{03} \hat{\cal Q}^{03} \nonumber\\
& & + 2 \, \sigma_1 \sigma_2 \hat{\cal Q}^{01} \hat{\cal Q}^{02} - 6 \, \sigma_2 \sigma_3 \hat{\cal Q}^{02} \hat{\cal Q}^{03}- 6 \, \sigma_1 \sigma_3 \hat{\cal Q}^{01} \hat{\cal Q}^{03} \biggr] 
\eea
\bea
\label{eq:5termsEx23}
& & \hskip-1.0cm V_{\hat{\cal P}\hat{\cal P}} =\frac{s}{4 \, {\cal V}_E^2} \biggl[\left(4 \sigma_2^2 - \sigma_1^2\right) \hat{\cal P}_0{}^1 \hat{\cal P}_0{}^1 + \left(\sigma_1^2 - \sigma_2^2\right) \hat{\cal P}_0{}^2 \hat{\cal P}_0{}^2  + \sigma_3^2 \hat{\cal P}_0{}^3 \hat{\cal P}_0{}^3 \nonumber\\
& & + 2 \, \sigma_1 \sigma_2 \hat{\cal P}_0{}^1 \hat{\cal P}_0{}^2 - 6 \, \sigma_2 \sigma_3 \hat{\cal P}_0{}^2 \hat{\cal P}_0{}^3- 6 \, \sigma_1 \sigma_3 \hat{\cal P}_0{}^1 \hat{\cal P}_0{}^3 \biggr] \hskip3.5cm \#(V_{\hat{\cal P}\hat{\cal P}})=16\nonumber\\
& & \hskip-0.0cm + \frac{s}{4 \, {\cal V}_E^2} \biggl[\left(4 \sigma_2^2 - \sigma_1^2\right) \hat{\cal P}^{01} \hat{\cal P}^{01} + \left(\sigma_1^2 - \sigma_2^2\right) \hat{\cal P}^{02} \hat{\cal P}^{02}  + \sigma_3^2 \hat{\cal P}^{03} \hat{\cal P}^{03} \nonumber\\
& & + 2 \, \sigma_1 \sigma_2 \hat{\cal P}^{01} \hat{\cal P}^{02} - 6 \, \sigma_2 \sigma_3 \hat{\cal P}^{02} \hat{\cal P}^{03}- 6 \, \sigma_1 \sigma_3 \hat{\cal P}^{01} \hat{\cal P}^{03} \biggr]
\eea
\bea
\label{eq:PQEx2}
& &  \hskip-1.0cm V_{\hat{\cal P}\hat{\cal Q}} =\frac{1}{4 \, {\cal V}_E^2} \times 2 \biggl[ (3\sigma_1^2-4\sigma_2^2)\, (\hat{\cal P}_0{}^1 \hat{\cal Q}^{01}-\hat{\cal Q}_0{}^1 \hat{\cal P}^{01})+ \sigma_3^2\, (\hat{\cal P}_0{}^3 \hat{\cal Q}^{03}-\hat{\cal Q}_0{}^3 \hat{\cal P}^{03})\nonumber\\
& & -(\sigma_1^2-3 \sigma_2^2)\, (\hat{\cal P}_0{}^2 \hat{\cal Q}^{02}-\hat{\cal Q}_0{}^2 \hat{\cal P}^{02})\nonumber\\
& &  + \sigma_1\, \sigma_2\, (\hat{\cal P}_0{}^2 \hat{\cal Q}^{01}-\hat{\cal Q}_0{}^2 \hat{\cal P}^{01} + \hat{\cal P}_0{}^1 \hat{\cal Q}^{02}-\hat{\cal Q}_0{}^1 \hat{\cal P}^{02}) \hskip3.3cm \#(V_{\hat{\cal P}\hat{\cal Q}})=22 \nonumber\\
& & + 5 \sigma_2\, \sigma_3\, (\hat{\cal P}_0{}^2 \hat{\cal Q}^{03}-\hat{\cal Q}_0{}^3 \hat{\cal P}^{02} + \hat{\cal P}_0{}^3 \hat{\cal Q}^{02}-\hat{\cal Q}_0{}^3 \hat{\cal P}^{02})\nonumber\\
& & +5 \sigma_1\, \sigma_3\, (\hat{\cal P}_0{}^1 \hat{\cal Q}^{03}-\hat{\cal Q}_0{}^3 \hat{\cal P}^{01}+\hat{\cal P}_0{}^3 \hat{\cal Q}^{01}-\hat{\cal Q}_0{}^1 \hat{\cal P}^{03}) \biggr] \,. 
\eea
So, we are done with rearranging the remaining 92 terms as well. Here we note that, the number of 960 terms in $F$-term scalar potential reduces to 382 terms for the case of switching off $P$-flux \cite{Shukla:2015hpa}. In the symplectic analysis of \cite{Shukla:2015hpa}, one can explicitly see how the volume moduli dependent factors of $V_{\hat{\cal Q}\hat{\cal Q}}$ (and so as for $V_{\hat{\cal P}\hat{\cal P}}$) can be read off from the following coefficient matrix depending on the K\"ahler moduli-space metric,
\bea
& & \hskip-1.5cm \left(\frac{4}{9}\, k_0^2 \tilde{\cal G}_{\alpha \beta} - 4 \sigma_\alpha \, \sigma_\beta\right) + \sigma_\alpha \, \sigma_\beta = \left(
 \begin{array}{ccc}
4 \sigma_2^2 -  \sigma_1^2& \sigma_1 \, \sigma_2 & -3 \sigma_1 \, \sigma_3\\
\sigma_1 \, \sigma_2 & \sigma_1^2 -  \sigma_2^2 & -3 \sigma_2 \, \sigma_3 \\
-3 \sigma_1 \, \sigma_3 & -3 \sigma_2 \, \sigma_3 & \sigma_3^2 \\
\end{array}
\right)
\eea
where $k_0 = 6 {\cal V}_E$ and $\tilde{\cal G}_{\alpha \beta}=\left(({\hat{d}^{-1}})_{\alpha}{}^{\alpha'}\, {\cal G}_{\alpha' \beta'}\, ({\hat{d}^{-1}})_{\beta}{}^{\beta'}\right)$ where ${\cal G}_{\alpha \beta} = -\frac{3}{2} \, \left( \frac{k_{\alpha \beta}}{k_0} - \frac{3}{2} \frac{k_\alpha \,k_\beta}{k_0^2}\right)$, and various non-zero intersection numbers for this orientifold setup are \cite{Robbins:2007yv}: $k_{311} = 1/2, k_{322} = -1, \hat{k}_{311} = -1, \hat{k}_{322} = -1/2$ along with $\hat{d}_\alpha{}^\beta = diag\{1/2, -1, 1/4\}$ and $d^a{}_b=diag\{-1,-1/2\}$.
\subsubsection*{D-term contributions}
The even-indexed flux orbits are,
\bea
& & {\cal R}_1 \equiv {\bf r}_1 = \frac{1 + |\tau|^2}{s} \, R_1, \, \, \hat{\mho}_{11} = \hat{\omega}_{11},\, \, \hat{\mho}_{21} = \hat{\omega}_{21}, \\
& & \hat{\mho}_{31} = \hat{\omega}_{31} - \left( Q_1{}^1 b_1 + Q_1{}^2 b_2 \right) - R_1 (2 b_1^2 + b_2^2 ) - \left( P_1{}^1 c_1 + P_1{}^2 c_2 \right) - R_1 (2 c_1^2 + c_2^2 )\nonumber
\eea
and similarly flux components with upper index $K = 1$ can be written. The $D$-term contributions to scalar potential results in a total of 210 terms which can be rearranged as,
\bea
\label{eq:5termsEx24}
& & \hskip-1.5cm V_{{\cal R}{\cal R}} = \frac{{\cal R}_1^2+({\cal R}^1)^2}{4\,f^2}  \hskip6.9cm \#(V_{{\cal R}{\cal R}}) = 12, \nonumber\\
& &  \hskip-1.5cm V_{{\hat{\cal \mho}}{\hat{\cal \mho}}} =  \frac{\left(t^\alpha \,\hat{{\cal \mho}}_{\alpha1}\right)^2+{(t^\alpha\, {\hat{\cal \mho}_\alpha{}^1})}^2 }{4\, \,{\cal V}_E^{2}}\, \hskip5cm \#( V_{{\hat{\cal \mho}}{\hat{\cal \mho}}}) =132, \nonumber\\
& & \hskip-1.5cm V_{{\cal R}{\hat{\cal \mho}}}=-2 \times \, \frac{{\cal R}_1 \left(t^\alpha \,\hat{{\cal \mho}}_{\alpha1}\right)+ {\cal R}^1 \, \left(t^\alpha_E\, {\hat{\cal \mho}_\alpha{}^1}\right)}{4 \,f\, {\cal V}_E} \hskip2.8cm \#(V_{{\cal R}{\hat{\cal \mho}}}) = 66\, , % \, \, \, {\cal R}_1 = r_1 \,, \, \, {\cal R}^1 = r^1. 
\eea
Thus, we have rewritten the four dimensional scalar potential using our new generalized flux orbits exemplifying the utility of the same. 

\section{Conclusions and discussions}
\label{sec_conclusions}
In this article, we have proposed some interesting combinations of generalized fluxes which help in rewriting the four dimensional effective scalar potential in a very compact manner. These peculiar flux combinations, which we call as {\it new generalized flux orbits}, respect the same strong/weak S-duality transformations as their respective old generalized flux orbits do. To be more specific, we have conjectured a generalized and S-duality invariant version of all the usual and (non-)geometric flux components, namely ($F, H, \omega, Q, P, R$) which are now denoted as (${\bf f}, {\bf h}, {\bf \omega}, {\bf q}, {\bf p}, {\bf r}$) respectively. These can be summarized in two classes; the ones which are counted by odd-index $k \in h^{2,1}_-(X)$ are collected as
\bea
\label{eq:orbits11A}
& &  {\bf h}_k = H_k + (\omega_{ak} \, {b}^a) + \hat{Q}^\alpha{}_k \, \left(\frac{1}{2}\, \hat{\kappa}_{\alpha a b} b^a b^b\right)  + \hat{P}^\alpha{}_k \, \left(\tilde{\rho}_\alpha -\frac{1}{2} \hat{\kappa}_{\alpha a b} c^a b^b\right) \nonumber\\
& &  {\bf f}_k = F_k + (\omega_{ak} \, {c}^a) - \hat{P}^\alpha{}_k \, \left(\frac{1}{2}\, \hat{\kappa}_{\alpha a b} c^a c^b\right) + \hat{Q}^\alpha{}_k \, \left(\tilde{\rho}_\alpha +\frac{1}{2} \hat{\kappa}_{\alpha a b} c^a b^b\right)\,, \nonumber\\
& &  {\bf h}^k = H^k + (\omega_{a}{}^{k} \, {b}^a) + \hat{Q}^{\alpha k} \, \left(\frac{1}{2}\, \hat{\kappa}_{\alpha a b} b^a b^b\right)  + \hat{P}^{\alpha k} \, \left(\tilde{\rho}_\alpha -\frac{1}{2} \hat{\kappa}_{\alpha a b} c^a b^b\right) \nonumber\\
& &  {\bf f}^k = F^k + (\omega_{a}{}^{k} \, {c}^a) - \hat{P}^{\alpha k} \, \left(\frac{1}{2}\, \hat{\kappa}_{\alpha a b} c^a c^b\right) + \hat{Q}^{\alpha k} \, \left(\tilde{\rho}_\alpha +\frac{1}{2} \hat{\kappa}_{\alpha a b} c^a b^b\right), \nonumber\\
& & {\bf \omega_{ak}} = \omega_{ak} + \hat{Q}^\alpha{}_k \, \left(\hat{\kappa}_{\alpha a b}\, b^b\right) -\, \hat{P}^\alpha{}_k \, \left(\hat{\kappa}_{\alpha a b}\, c^b\right) \\
& & {\bf \omega_{a}{}^{k}}= \omega_{a}{}^{k} + \hat{Q}^{\alpha k} \, \left(\hat{\kappa}_{\alpha a b}\, b^b\right)-\hat{P}^{\alpha k} \, \left(\hat{\kappa}_{\alpha a b}\, c^b\right)\nonumber\\
& & {\bf \hat{q}^{\alpha}{}_{k}} = \hat{Q}^{\alpha}{}_{k} , \quad {\bf \hat{q}^{\alpha k}} = \hat{Q}^{\alpha k}, \quad \quad \quad {\bf \hat{p}^{\alpha}{}_{k}} = \hat{P}^{\alpha}{}_{k}, \quad {\bf {\hat{p}}^{\alpha k}} = {\hat{P}}^{\alpha k} \, .\nonumber
\eea
while those counted by even-index $K \in h^{2,1}_+(X)$ are given as under,
\bea
\label{eq:orbits11B}
& & {\bf \hat{\omega}_{\alpha K}} = \hat{\omega}_{\alpha K}\, - (d^{-1})_b{}^a  \, Q^{b}{}_{K} \, \, \left(\hat{k}_{\alpha a c} \, b^c \right) + f^{-1} \, \, R_K \, \left(\frac{1}{2}\hat{k}_{\alpha a b} \, b^a \,b^b\right)\nonumber\\
& & \hskip2.4cm + (d^{-1})_b{}^a  \, P^{b}{}_{K} \, \, \left(\hat{k}_{\alpha a c} \, c^c \right) + f^{-1} \, \, R_K \, \left(\frac{1}{2}\hat{k}_{\alpha a b} \, c^a \,c^b\right)\nonumber\\
& & {\bf \hat{\omega}_{\alpha}{}^{K}} =\hat{\omega}_{\alpha}{}^{K}\, - (d^{-1})_b{}^a \, \, Q^{b K} \, \left(\hat{k}_{\alpha a c} \, b^c\right)+ f^{-1} \, \, R^K \, \left(\frac{1}{2}\hat{k}_{\alpha a b} \, b^a \,b^b\right)\nonumber\\
& & \hskip2.4cm + (d^{-1})_b{}^a \, \, P^{b K} \, \left(\hat{k}_{\alpha a c} \, c^c\right)+ f^{-1} \, \, R^K \, \left(\frac{1}{2}\hat{k}_{\alpha a b} \, c^a \,c^b\right)\nonumber\\
& & {\bf q}^{a{}K} ={Q}^{a{}K} + f^{-1} \,\, d_b{}^a  \, (R^K \bullet \, b^b)\,, \quad \quad {\bf q}^{a}{}_{K} = {Q}^{a}{}_{K} + f^{-1} \,\,d_b{}^a \, (R_K \bullet b^b) \\
& & {\bf p}^{a{}K} ={P}^{a{}K} - f^{-1} \,\, d_b{}^a  \, (R^K \bullet \, c^b)\,, \quad \quad {\bf p}^{a}{}_{K} = {P}^{a}{}_{K} - f^{-1} \,\,d_b{}^a \, (R_K \bullet c^b) \nonumber\\
& & {\bf r_K}= \frac{(1+|\tau|^2)}{s}\, R_K\, , \quad \quad \quad  {\bf r^K}= \frac{(1+|\tau|^2)}{s}\, R^K\, . \nonumber
\eea
These flux orbits consistently reproduce the results of \cite{Taylor:1999ii, Blumenhagen:2003vr, Gao:2015nra, Blumenhagen:2013hva, Shukla:2015bca}. Moreover, from these representations of new generalized flux combinations as summarized in eqns. (\ref{eq:orbits11A})-(\ref{eq:orbits11B}), using eqn. (\ref{eq:S-duality}) it is obvious that the orbits transform under S-duality: $\tau \to -\frac{1}{\tau}$ as,
\bea
\label{eq:orbits11C}
& & \hskip0.5cm  {\bf f^k} \to  -{\bf h^k}, \quad {\bf h^k} \to {\bf f^k}, \quad  {\bf {\cal \omega}_{a}{}^{k}} \to {\bf {\cal \omega}_{a}{}^{k}},  \quad {\bf \hat{q}^{\alpha k}} \to - {\bf \hat{p}^{\alpha k}},   \quad  {\bf \hat{p}^{\alpha k}} \to  {\bf \hat{q}^{\alpha k}} \\
& & \hskip0.5cm {\bf f_k} \to  -{\bf h_k}, \quad {\bf h_k} \to {\bf f_k}, \quad  {\bf {\cal \omega}_{ak}} \to {\bf {\cal \omega}_{a k}},  \quad {\bf \hat{q}^{\alpha}{}_{k}} \to - {\bf \hat{p}^{\alpha}{}_{k}},   \quad  {\bf \hat{p}^{\alpha}{}_{ k}} \to  {\bf \hat{q}^{\alpha}{}_{k}} \nonumber\\
%& & \hskip-2.2cm \tau \to -\frac{1}{\tau}  \Longrightarrow \nonumber\\
& & \hskip0.5cm {\bf \hat{{\mathbb\omega}}_{\alpha}{}^{K}} \to  {\bf \hat{{\mathbb\omega}}_{\alpha}{}^{K}}, \quad {\bf {q}^{a K}} \to - {\bf {p}^{a K}},  \quad  {\bf {p}^{a K}} \to  {\bf {q}^{a K}}, \quad {\bf r}^K \to {\bf r}^K \,  \nonumber\\
& & \hskip0.5cm {\bf \hat{{\mathbb\omega}}_{\alpha K}} \to  {\bf \hat{{\mathbb\omega}}_{\alpha K}}, \quad {\bf {q}^{a}{}_{K}} \to - {\bf {p}^{a}{}_{ K}},  \quad  {\bf {p}^{a}{}_{ K}} \to  {\bf {q}^{a}{}_{ K}}, \quad {\bf r}_K \to {\bf r}_K \, . \nonumber
\eea
By respecting the same transformations as earlier in eqn. (\ref{eq:S-duality}), this analysis shows that the new generalized flux orbits provide a completion of the usual generalized flux orbits under strong/weak duality. Moreover after examining the transformation of new generalized flux orbits under $\tau \to \tau +1$ by using eqn. (\ref{eq:S-duality1}) we find that 
\bea
\label{eq:orbits11D}
& & \hskip-0.5cm  {\bf f^k} \to {\bf f^k} -{\bf h^k}, \quad {\bf h^k} \to {\bf h^k}, \quad  {\bf {\cal \omega}_{a}{}^{k}} \to {\bf {\cal \omega}_{a}{}^{k}},  \quad {\bf \hat{q}^{\alpha k}} \to {\bf \hat{q}^{\alpha k}} - {\bf \hat{p}^{\alpha k}},   \quad  {\bf \hat{p}^{\alpha k}} \to  {\bf \hat{p}^{\alpha k}} \\
& & \hskip-0.5cm {\bf f_k} \to {\bf f_k} -{\bf h_k}, \quad {\bf h_k} \to {\bf h_k}, \quad  {\bf {\cal \omega}_{ak}} \to {\bf {\cal \omega}_{a k}},  \quad {\bf \hat{q}^{\alpha}{}_{k}} \to {\bf \hat{q}^{\alpha}{}_{k}} - {\bf \hat{p}^{\alpha}{}_{k}},   \quad  {\bf \hat{p}^{\alpha}{}_{ k}} \to  {\bf \hat{p}^{\alpha}{}_{k}} \nonumber
\eea
Subsequently, it is interesting to observe that the orbit combinations, 
\begin{eqnarray*}
\label{eq:orbitsB1}
& \hskip-1.5cm {\cal H}_k =  {\bf h}_k~, \, \, \, \, \, &\hat{\cal Q}^{\alpha}{}_{k} ={ \bf{\hat{q}^{\alpha}{}_{k} }} + C_0 \, {\bf \hat{p}^{\alpha}{}_{k}} ~, \nonumber\\
& {\cal F}_k=   {\bf f}_k + C_0 \, ~{\bf h}_k~, \, \, \, \, \, &\hat{\cal P}^{\alpha}{}_{k} = {\bf \hat{p}^{\alpha}{}_{k}}~; \nonumber\\
& \hskip-1.5cm {\cal H}^k =  {\bf h}^k~, \, \, \, \, \, &{\hat{\cal Q}}^{\alpha k} = {\bf \hat{q}^{\alpha k}} + C_0 \, {\bf \hat{p}^{\alpha k}} ~, \nonumber\\
& {\cal F}^k=   {\bf f}^k + C_0 \, ~{\bf h}^k~, \, \, \, \, \, &{\hat{\cal P}}^{\alpha k} = {\bf {\hat{p}}^{\alpha k}}~, \nonumber
\end{eqnarray*}
are invariant under $\tau \to \tau +1$. Recall that the final rearrangements of $F$-term contributions (for Model A and Model B) are written out using these (${\cal H}_k, {\cal F}_k$ etc.) flux orbits, and so the same ensure the invariance of total $F$-term scalar potential under $\tau \to -\frac{1}{\tau}$ and $\tau \to \tau + 1$. This has been well anticipated as argued  earlier in the end of section 2 that effectively $K \to K$ and $W \to W$ under $\tau \to \tau +1$. 

However, under $\tau \to \tau + 1$, the situation for $D$-term contributions does not appear to be as clean as it was for $F$-terms, and we find that new geometric flux orbits $({\bf \hat{{\mathbb\omega}}_{\alpha}{}^{K}}, {\bf \hat{{\mathbb\omega}}_{\alpha K}})$ remain invariant only in the absence of non-geometric $R$-flux. Moreover, the orbits of R-flux $({\bf r_K, r^K})$ are also not invariant under $\tau \to \tau +1$. Nevertheless, we find that at least in the absence of $R$-flux (which still results in non-trivial D-terms with other fluxes), these problems disappear, and similar transformations to those of odd-indexed flux orbits (as in eqn. (\ref{eq:orbits11D})) holds,
\bea
%& & \hskip-2.2cm   \Longrightarrow \nonumber\\
& & \hskip0.5cm {\bf \hat{{\mathbb\omega}}_{\alpha}{}^{K}} \to  {\bf \hat{{\mathbb\omega}}_{\alpha}{}^{K}}, \quad {\bf {q}^{a K}} \to {\bf {q}^{a K}}- {\bf {p}^{a K}},  \quad  {\bf {p}^{a K}} \to  {\bf {p}^{a K}} \,  \nonumber\\
& & \hskip -2.5cm {\rm For \, \,} R = 0 \\
& & \hskip0.5cm {\bf \hat{{\mathbb\omega}}_{\alpha K}} \to  {\bf \hat{{\mathbb\omega}}_{\alpha K}}, \quad {\bf {q}^{a}{}_{K}} \to {\bf {q}^{a}{}_{K}} - {\bf {p}^{a}{}_{ K}},  \quad  {\bf {p}^{a}{}_{ K}} \to  {\bf {p}^{a}{}_{ K}} \, . \nonumber
\eea
and subsequently the doublets $({\cal Q}^a_K := {\bf q^a_K} + C_0 \, {\bf p^a_K}, \, {\bf p^a_K})$ and $({\cal Q}^{aK} := {\bf q^{aK}} + C_0 \, {\bf p^{aK}},\, {\bf p^{aK}})$ are also invariant under $\tau \to \tau +1$ if there is no $R$-flux present. Note that, the nature of $R$-flux remains mysterious in many aspects, and despite being allowed by the orietifold projection, what kind of fluxes can be consistently/simultaneously turned-on, still remains an open issue. On these lines, this analysis may be considered as a hint towards ruling out the possibility of turning on $R$-flux in a completely modular invariant framework of type IIB  superstring theory. However, it will be interesting to check by some more justified approach (e.g. via carefully analyzing various Jacobi identities) if $R$-fluxes could indeed not be allowed within a type IIB compactification framework with involutions leading to $O3/O7$-planes. Nevertheless, we note again that the transformations under strong/weak duality, $\tau \to -\frac{1}{\tau}$ are respected by our new generalized flux orbits even in the presence of $R$-flux, while satisfying those of $\tau \to \tau+1$ demands to switch off the non-geometric $R$-flux.  %The additional axionic shift inside $\tau \to \tau +1$ creates the subtlety which requires more investigations. In addition, the possibility of another set of modular completed  expressions, different from those of ours, though unlikely, but is not ruled out. 

In the second half of this article, by giving two explicit examples with frozen complex structure moduli, we have illustrated how the proposed new generalized flux orbits help in rewriting the full four dimensional effective scalar potential in a very compact manner which could be useful for exploring the 10D non-geometric action on the lines of \cite{Blumenhagen:2013hva,Gao:2015nra}. Moreover, the compact representation of scalar potential may ease the moduli stabilization procedure and could subsequently help in studying the subsequent phenomenological applications. It is worth to mention that after this proposal for {\it new} generalized flux orbits was made, the same have already been tested  to be useful for understanding the full scalar potential from ten-dimensional perspective in the absence of S-dual P-flux \cite{Shukla:2015bca, Shukla:2015hpa}. Moreover, these flux orbits are also closely related to the flux combinations utilized while studying flux formulation of Double Field Theory (DFT) reduction on Calabi Yau orientifolds \cite{Blumenhagen:2015lta}. Now, it would be interesting to check our new orbit expressions of various fluxes via extending the symplectic formulation of \cite{Shukla:2015hpa} as well as DFT reduction of \cite{Blumenhagen:2015lta} to include the S-dual P-fluxes, and we hope to get back to this aspect in future.

\section*{Acknowledgments}
I am thankful to Ralph Blumenhagen for useful discussions and continuous encouragements. Moreover, I also thank Anamaria Font, Xin Gao, Daniela Herschmann, Oscar Loaiza-Brito and Erik Plauschinn for useful discussions. This work was supported by the Compagnia di San Paolo contract ``Modern Application of String Theory'' (MAST) TO-Call3-2012-0088.

\appendix
\section{Appendix}
To illustrate how the counting of terms is done for various collection of pieces in Model A and Model B, here we present the expanded version of the first piece $V_{{\cal H}{\cal H}}$ given in eqn. (\ref{eq:5termsEx2}). In the usual flux orbits, this piece has 72 terms which are compactly written into just two-terms by using our new generalized flux orbits.
\bea
\label{eq:append1}
& & \hskip-0.8cm V_{{\cal H}{\cal H}}=  \frac{s }{4 \, {\cal V}_E^2} \biggl[H_0^2 +\left(H^0\right)^2+\left(\omega _{10}\right)^2 \, b_1^2  +\left(\omega _1{}^0\right){}^2 \, b_1^2+  \left(\omega _{20}\right)^2 \, b_2^2 \, +b_2^2\, \left(\omega _2{}^0\right){}^2  +2 \, b_2 \, H_0 \,\omega_{20} \nonumber\\
& & +b_2 \, \kappa_{311} \, \omega_{20}\, \hat{Q}_0{}^3 \, b_1^2  +\kappa_{311} \, \rho_1 \, \hat{P}_0{}^1 \, \hat{Q}_0{}^3 \, b_1^2 +\kappa_{311} \, \rho_2 \, \hat{P}_0{}^2 \, \hat{Q}_0{}^3 \, b_1^2+\kappa_{311} \, \rho_3 \, \hat{P}_0{}^3 \, \hat{Q}_0{}^3 \, b_1^2\nonumber\\
& & +\kappa_{311} \, H^0 \, \hat{Q}^{30}\, b_1^2 + \kappa_{311} \, \omega_2{}^0 \, \hat{Q}^{30} \, b_1^2\, b_2 +\kappa_{311} \, \rho_1 \, \hat{P}^{10}\, \hat{Q}^{03} \,b_1^2+\kappa_{311} \, \rho_2 \, \hat{P}^{20}\, \hat{Q}^{30}\, b_1^2 \nonumber\\
& & +\kappa_{311} \, \rho_3 \, \hat{P}^{30}\, \hat{Q}^{30}\, b_1^2+2\, H_0 \, \omega_{10}\, b_1+ 2\,  b_2 \, \omega_{10}\, \omega_{20}\,  b_1+2 \, \rho_1 \, \omega_{10}\, \hat{P}_0{}^1 \, b_1+2 \, \rho_2 \, \omega_{10}\, \hat{P}_0{}^2 \, b_1\nonumber\\
& & +2 \rho_3 \, \omega _{10}\, \hat{P}_0{}^3 \, b_1+b_2^2 \, \kappa_{322} \, \omega_{10}\, \hat{Q}_0{}^3 \, b_1+2 \, H^0 \, \omega _1{}^0 \, b_1+2\, b_2 \,\omega_1{}^0 \, \omega _2{}^0 \, b_1+2 \rho_1 \, \omega _1{}^0 \, \hat{P}^{10}\, b_1\nonumber\\
& & + 2\, b_2 \, \rho_1 \, \omega_{20}\,  \hat{P}_0{}^1 +2 \, H_0\, \rho_2\, \hat{P}_0{}^2 + 2\, b_2 \, \rho_2 \, \omega_{20} \, \hat{P}_0{}^2 + 2\, \rho_1\, \rho_2\, \hat{P}_0{}^1\, \hat{P}_0{}^2 + 2\,  H_0 \, \rho_3 \, \hat{P}_0{}^3 \nonumber\\
& & + 2\,  b_2\,  \rho_3 \, \omega_{20}\, \hat{P}_0{}^3 +2 \rho_1 \, \rho_3\, \hat{P}_0{}^1 \, \hat{P}_0{}^3+2 \, \rho_2 \, \rho_3 \, \hat{P}_0{}^2 \, \hat{P}_0{}^3+b_2^2 \, H_0 \, \kappa_{322} \, \hat{Q}_0{}^3+ b_2^3 \, \kappa_{322} \, \omega_{20} \, \hat{Q}_0{}^3 \nonumber\\
& & +b_2^2 \, \kappa_{322} \, \rho_1\, \hat{P}_0{}^1 \, \hat{Q}_0{}^3 + b_2^2 \, \kappa_{322}\, \rho_2 \, \hat{P}_0{}^2 \, \hat{Q}_0{}^3 + b_2^2 \, \kappa_{322} \, \rho_3 \, \hat{P}_0{}^3 \, \hat{Q}_0{}^3+2 \,b_2\, H^0\, \omega_2{}^0+2 \, \rho_1 \, H^0 \,  \hat{P}^{10} \nonumber\\
& & +2 \, b_2 \, \rho_1 \, \omega_2{}^0\,  \hat{P}^{10} + 2 \,  \rho_2 \, H^0\,  \hat{P}^{20} + 2\, b_2 \, \rho_2 \, \omega_2{}^0 \, \hat{P}^{20} + 2\,  \rho_1 \, \rho_2 \, \hat{P}^{10} \, \hat{P}^{20} + 2 \, \rho_3\, H^0\, \hat{P}^{30} \nonumber\\
& & +2 b_2 \, \rho_3 \, \omega_2{}^0 \, \hat{P}^{30} + 2\, \rho_1\, \rho_3\, \hat{P}^{10}\, \hat{P}^{30} + 2\, \rho_2 \, \rho_3 \, \hat{P}^{20} \, \hat{P}^{30} + b_2^2 \, \kappa_{322}\,  H^0 \, \hat{Q}^{30} +b_2^3 \, \kappa_{322} \, \omega_2{}^0 \, \hat{Q}^{30}\nonumber\\
& &  + b_2^2 \, \kappa_{322} \, \rho_1 \, \hat{P}^{10}\,  \hat{Q}^{30} + b_2^2 \, \kappa _{322} \, \rho_2 \, \hat{P}^{20}\, \hat{Q}^{30} + b_2^2 \, \kappa_{322} \, \rho_3 \, \hat{P}^{30}\, \hat{Q}^{30} + 2\,  H_0 \, \rho_1 \, \hat{P}_0{}^1 +2 \, \rho_2 \, \omega_1{}^0 \, \hat{P}^{20}\, b_1 \nonumber\\
& &  +\kappa_{311} \, \omega_{10} \, \hat{Q}_0{}^3 \, b_1^3 +\kappa_{311} \, \omega _1{}^0 \, \hat{Q}^{30}\, b_1^3 + \kappa_{322} \, \omega _1{}^0 \, \hat{Q}^{30}\, b_1\, b_2^2 + H_0 \, \kappa_{311} \, {\hat Q}_0{}^3 \, b_1^2 +2\, \rho_3 \, \omega_1{}^0 \, \hat{P}^{30}\, b_1 \nonumber\\
& & +\rho_1^2 \, \left(\hat{P}_0{}^1\right){}^2+\rho_2^2 \left(\hat{P}_0{}^2\right){}^2+\rho_3^2 \left(\hat{P}_0{}^3\right){}^2  +\rho_1^2 \left(\hat{P}^{10}\right)^2 +\rho_2^2 \left(\hat{P}^{20}\right)^2 +\rho_3^2 \left(\hat{P}^{30}\right)^2 \nonumber\\
& &  +\frac{1}{4} \, \kappa_{322}^2 \, b_2^4 \, \left(\hat{Q}_0{}^3\right){}^2+\frac{1}{4} \, b_2^4 \, \kappa_{322}^2 \, \left(\hat{Q}^{30}\right)^2 + \frac{1}{4} \kappa_{311}^2 \left(\hat{Q}_0{}^3\right){}^2 b_1^4 +\frac{1}{4} \kappa_{311}^2 \left(\hat{Q}^{30}\right)^2 b_1^4 \nonumber\\
& & +\frac{1}{2}  \kappa_{311} \, \kappa_{322} \,\left(\hat{Q}_0{}^3\right){}^2 \, b_1^2 \, b_2^2  +\frac{1}{2}  \kappa_{311} \, \kappa _{322} \left(\hat{Q}^{30}\right)^2 \, b_1^2 \, b_2^2  \biggr]\nonumber\\
& & \equiv  \frac{s }{4 \, {\cal V}_E^2}\, \biggl[{\cal H}_0^2 +\, {({\cal H}^0)}^2 \biggr] 
\eea
This shows how crucial are our new generalized flux orbits as the same help in rewriting 72 terms of F-term scalar potential into just two terms ! Other (more complicated) pieces of collections given in eqns. (\ref{eq:5termsEx2})-(\ref{eq:PQEx2}}) have been invoked in a similar fashion via the three iterative steps as we have earlier mentioned.

\newpage
%\bibliography{references}
\bibliographystyle{utphys}
\bibliography{SdualReference}

\end{document}